\documentclass[fontsize=10pt,a4paper]{scrartcl}
\usepackage[a4paper, top=4cm, bottom=4cm]{geometry}
\usepackage[utf8x]{inputenc}
\usepackage[T1]{fontenc}
\usepackage{tikz}
\usetikzlibrary{matrix,arrows}
\usepackage{enumitem}
\usepackage{mwe}
\usepackage{hyperref}
\usepackage{euscript}
\usepackage{amsmath}
\usepackage{amsthm}
\usepackage{amssymb}
\usepackage{amsfonts}
\usepackage{mathrsfs}
\usepackage{bbm}
\usepackage{abstract}
\usepackage{graphicx}
\usepackage{placeins}
\usepackage{subfigure}
\usepackage{nicefrac}
\usepackage{booktabs}
\usepackage{bookmark}
\usepackage{multirow}
\usepackage{dsfont}
\usepackage{longtable}
\usepackage{braket}
\usepackage{tikz}
\usepackage{scalerel}
\usepackage{listings}
\usepackage{color}
\usepackage[all]{xy}
\usepackage{tensor}
\usepackage{faktor}
\usepackage{lipsum}
\usepackage{comment}
\usepackage{cite}
\usepackage{arydshln}

\theoremstyle{definition}
\newtheorem{definition}{Definition}[section] 
 
\theoremstyle{definition}
\newtheorem{remark}[definition]{Remark}

\theoremstyle{definition}

\theoremstyle{plain}

\newtheorem{prop}[definition]{Proposition}

\newcommand{\Osp}[2]{\mathrm{OSp}(#1|#2)} 
\newcommand{\UOsp}[2]{\mathrm{UOSp}(#1|#2)} 
\newcommand{\SLC}{\mathrm{SL}(2,\C)}

\newcommand{\C}{\mathbb{C}}

\newcommand{\calN}{\mathcal{N}}

\begin{document}
\title{\LARGE Towards black hole entropy in chiral loop quantum supergravity}
\author{
  \large Konstantin Eder\thanks{konstantin.eder@gravity.fau.de}\, and Hanno Sahlmann\thanks{hanno.sahlmann@gravity.fau.de} \\
	\large Institute for Quantum Gravity (IQG)\\
		\large Friedrich-Alexander-Universität Erlangen-Nürnberg (FAU)
}
\maketitle

\begin{abstract}
Recently, many geometric aspects of $\mathcal{N}$-extended AdS supergravity in chiral variables have been encountered and clarified. In particular, if the theory is supposed to be invariant under SUSY transformations also on boundaries, the boundary term has to be the action of a $\Osp{\mathcal{N}}{2}_{\C}$ super Chern-Simons theory, and particular boundary conditions must be met. 

Based on this, we propose a way to calculate an entropy $S$ for surfaces, presumably including black hole horizons, in the supersymmetric version of loop quantum gravity for the minimal case $\mathcal{N}=1$. It proceeds in analogy to the non-supersymmetric theory, by calculating dimensions of quantum state spaces of the super Chern-Simons theory with punctures, for fixed quantum (super) area of the surface. We find $S = a_H/4$ for large areas and determine the subleading correction.

Due to the non-compactness of $\Osp{1}{2}_{\C}$ and the corresponding difficulties with the Chern-Simons quantum theory, we use analytic continuation from the Verlinde formula for a compact real form, $\UOsp{1}{2}$, in analogy to work by Noui et al. This also entails studying some properties of $\Osp{1}{2}_{\C}$ representations that we have not found elsewhere in the literature.   
\end{abstract}

\newpage

\section{Introduction}\label{superCartan geometry}
Since there are indications that horizons can meaningfully be assigned a thermodynamic entropy \cite{Bardeen:1973gs,Bekenstein:1973ur,Hawking:1975vcx}, the challenge is to explain it as the von Neumann-type entropy of a quantum description of the black hole. It has met with some measure of success in string theory (entropy of BPS black holes, for example \cite{Strominger:1996sh,Behrndt:1996jn}) and loop quantum gravity (entropy of isolated horizons, for example \cite{Smolin:1995vq,Rovelli:1996dv,Ashtekar:1997yu,Kaul:1998xv,Domagala:2004jt,Meissner:2004ju,Engle:2009vc,Agullo:2009eq}). Interestingly, the two approaches are very different in nature and the results concern disjoint families of black holes. 

The current work starts to bridge this gap, by considering the entropy of certain surfaces in supergravity, quantized with methods from LQG. The theory we consider is $\calN=1, D=4$ supergravity. The idea is to calculate the entropy as the log of the size of the space of quantum states of a super Chern-Simons theory. This theory is constrained by the surface area, and hence the entropy becomes area dependent.
But since the structure group is non-compact, the quantum theory for this super Chern-Simons theory is not directly accessible. Rather, we start from the state counting for a Chern-Simons theory with a compact structure group, and analytically continue the result in a particular way, following closely the procedure for the non-supersymmetric theory set out in \cite{Frodden:2012dq,Achour:2014eqa}, see also \cite{Han:2014xna}.  
We find that
\begin{equation}
\label{eq:1}
    S=\frac{a_H}{4l_p^2}+\mathcal{O}(\sqrt{a_H}/l_p)
\end{equation}
where $a_H$ is the diffeomorphism and gauge invariant measure of area in the supergeometric setting. 

The present work is not the first that is considering black hole entropy in supergravity from a loop quantum gravity perspective. We are using variables that were first proposed in \cite{Fulop:1993wi} and whose geometric meaning was recently clarified in \cite{Eder:2020erq,Eder:2021rgt,Eder:2022}. A super Chern-Simons theory as a source of entropy was first considered in \cite{Smolin:1995vq}. Supergravity with loop quantum gravity methods has also been considered in \cite{Ling:1999gn}, and using different variables in \cite{Bodendorfer:2011hs,Bodendorfer:2011pb,Bodendorfer:2011pc}. Our treatment follows \cite{Fulop:1993wi,Smolin:1995vq,Ling:1999gn,Gambini:1995db} in keeping supersymmetry manifest, but it goes beyond it and the other works by making use of a detailed geometric analysis of the super Ashtekar connection and corresponding boundary conditions. It is also the first that is based on a detailed state counting in super Chern-Simons theory, as far as we know. Moreover, this is the first time that the Bekenstein-Hawking area law is derived and verified within the supersymmetric setting.

Let us explain the setup and strategy. The quantum theory is obtained from a canonical formulation of $\calN=1, D=4$ supergravity in terms of a supersymmetric generalization \cite{Fulop:1993wi,Eder:2020erq,Eder:2021rgt,Eder:2022} of the (chiral) Ashtekar connection $\mathcal{A}^+$ which can be obtained from a Holst modification of the McDowell Mansouri action \cite{Eder:2021rgt,Eder:2022}. The structure group in this formulation is $\Osp{1}{2}$.
We consider this theory in the presence of a causal boundary of spacetime, playing the role of the horizon. The requirement of local supersymmetry also on the boundary \emph{uniquely} fixes a supersymmetric boundary term 
\begin{equation}
    S_{\text{bdy}}(\mathcal{A}^+)=\frac{k}{4\pi}\int_{H}{\braket{\mathcal{A}^+\wedge\mathrm{d}\mathcal{A}^++\frac{1}{3}\mathcal{A}^+\wedge[\mathcal{A}^+\wedge\mathcal{A}^+]}}
\end{equation}
that is given by an $\Osp{1}{2}$ Chern-Simons theory and boundary conditions
\begin{equation}
    \underset{\raisebox{1pt}{$\Leftarrow$}}{F(\mathcal{A}^{+})}\propto \underset{\raisebox{1pt}{$\Leftarrow$}}{\mathcal{E}}
\end{equation}
linking curvature and super-electric field on the boundary. As in the non-supersymmetric case, the idea is to quantize bulk and boundary separately and couple them via the boundary condition. In this picture, field excitations in the bulk couple to Chern-Simons defects in the boundary theory. To flesh out this picture, we provide a sketch of the bulk quantum theory (in fact, for $\mathcal{N}=1$ and $2$), including the definition of the graded holonomy-flux algebra, supersymmetric generalizations of spin networks, and the supersymmetric area operator. 

$\Osp{1}{2}$ is non-compact, however, so there are fundamental technical problems in defining the Hilbert space for the bulk theory. This is very similar for the original Ashtekar variables with structure group $\SLC$. For the entropy calculation, we therefore start from the Chern-Simons theory of a compact real form of this group, $\UOsp{1}{2}$ and use analytical continuation in the corresponding Verlinde type formula that is counting its states. This procedure is a generalization of that employed in \cite{Frodden:2012dq,Achour:2014eqa}. 

We finally note that the present calculation is different from the string theory one in some respects. 
Our calculation here seems to apply to a large class of surfaces that carry local supersymmetry, whereas in string theory more restrictive class of surfaces corresponding to BPS black holes are considered. Fermionic degrees of freedom play no direct role in that calculation. This is in contrast to our situation in which fermionic degrees of freedom are taken into account in the entropy calculation. That fermionic degrees of freedom may in fact lead to interesting consequences in the context of supersymmetric black holes has been observed in \cite{Aichelburg:1983ux}, where it has been shown that the supersymmetric black holes can carry a nontrivial (fermionic) supercharge which may also contribute to the first law of black hole mechanics.

Let us finish this introduction with a summary of the structure of the work. In section \ref{sec:Review} we briefly review the classical setting, including the supergravity action we use, boundary terms and boundary conditions, the super-Ashtekar connection and the resulting symplectic structure. 
In section \ref{sec:qt}, we sketch the quantum theory of the bulk, in particular the precise definition of the graded holonomy-flux algebra, the possibilities and issues in connection with the bulk Hilbert space, and the action of the super area operator. We also discuss the boundary Chern-Simons theory and its coupling to the bulk theory. 
In section \ref{sec:ospreps}, we define and discuss a continuous family of representations of $\UOsp{1}{2}$ that is relevant for the entropy calculation, as well as interesting in its own right. 
\ref{sec:EntropyBH} contains the determination of the size of the state space of the $\mathrm{UOSp}(1|2)_k$ Chern-Simons theory in the limit of large $k$ and its analytic continuation and asymptotic analysis for the physically relevant case. The entropy formula \eqref{eq:1} is established in that section. 
The article ends with a discussion of the results and open questions. The appendix deals with super Chern-Simons theory (appendix \ref{Appendix:Chern-Simons}) as well as the relevant supergroups (appendix \ref{Appendix:Supergroups}).

\section{Review: The Holst-MacDowell-Mansouri action of chiral supergravity}\label{sec:Review}
In this section, let us briefly review the Cartan geometric description of pure AdS Holst-super-gravity with $\mathcal{N}$-extended supersymmetry with $\mathcal{N}=1,2$. For more details, we refer to \cite{Eder:2022,Eder:2021rgt,Eder:2020erq} (see also \cite{Andrianopoli:2014aqa,Andrianopoli:2020zbl} using standard variables).\\
Pure AdS (Holst-)supergravity can be described in terms of a super Cartan geometry modeled on the super Klein geometry $(\mathrm{OSp}(\mathcal{N}|4),\mathrm{Spin}^+(1,3)\times\mathrm{SO}(\mathcal{N}))$ with super Cartan connection 
\begin{equation}
\mathcal{A}=e^IP_I+\frac{1}{2}\omega^{IJ}M_{IJ}+\frac{1}{2}\hat{A}_{rs}T^{rs}+\Psi_r^{\alpha}Q_{\alpha}^r
\label{D.eq:5.1.0}
\end{equation}
This connection can be used in order to formulate a Yang-Mills-type action principle for Holst-supergravity. To this end, one introduces a $\beta$-deformed inner product $\braket{\cdot\wedge\cdot}_{\beta}$ on $\mathfrak{g}\equiv\mathfrak{osp}(\mathcal{N}|4)$-valued differential forms on the underlying spacetime manifold $M$ with $\beta$ the Barbero-Immirzi parameter via
\begin{align}
    \braket{\cdot\wedge\cdot}_{\beta}:\,\Omega^2(M,\mathfrak{g})\times\Omega^2(M,\mathfrak{g})&\rightarrow\Omega^4(M)\\
    (\omega,\eta)&\mapsto\mathrm{str}(\omega\wedge\mathbf{P}_{\beta}\eta)
\end{align}
with ``$\mathrm{str}$'' denoting the $\mathrm{Ad}$-invariant supertrace on $\mathfrak{g}$ and $\mathbf{P}_{\beta}$ a $\beta$-dependent operator on $\Omega^2(M,\mathfrak{g})$ (the precise form of this operator does not matter in what follows; for more details see \cite{Eder:2022,Eder:2021rgt}). Using this inner product, the so-called Holst-MacDowell-Mansouri action of $\mathcal{N}$-extended pure Ads Holst-supergravity takes the form
\begin{equation}
S^{\beta}_{\text{H-MM}}(\mathcal{A})=\frac{L^2}{\kappa}\int_{M}{\braket{F(\mathcal{A})\wedge F(\mathcal{A})}_{\beta}}
\label{D.eq:6.4}
\end{equation}
with $F(\mathcal{A})$ the Cartan curvature of the super Cartan connection $\mathcal{A}$.

In the chiral limit of the theory corresponding to an imaginary $\beta=-i$, the Holst-MacDowell-Mansouri action \eqref{D.eq:6.4} becomes manifestly invariant under an enlarged $\mathrm{Osp}(\mathcal{N}|2)_{\mathbb{C}}$-gauge symmetry. In fact, in this limit, it follows that the operator $\mathbf{P}_{-i}$ decomposes as $\mathbf{P}_{-i}=\tilde{\mathbf{P}}_{-i}\circ\mathbf{P}^{\mathfrak{osp}(\mathcal{N}|2)}$ with $\mathbf{P}^{\mathfrak{osp}(\mathcal{N}|2)}:\,\mathfrak{osp}(\mathcal{N}|4)\rightarrow\mathfrak{osp}(\mathcal{N}|2)_{\mathbb{C}}$ the projection operator onto the (complexified) chiral sub superalgebra $\mathfrak{osp}(\mathcal{N}|2)_{\mathbb{C}}$ of $\mathfrak{g}$. Applying this operator on the super Cartan connection \eqref{D.eq:5.1.0} this yields the super Asthekar connection 
\begin{equation}
\mathcal{A}^{+}:=\mathbf{P}^{\mathfrak{osp}(\mathcal{N}|2)}\mathcal{A}=A^{+ i}T_i^{+}+\psi^A_r Q_A^r+\frac{1}{2}\hat{A}_{rs}T^{rs}
\label{D.eq:SuperAshtekarGeneral}
\end{equation}
Using this connection, it then follows that the Holst-MacDowell-Mansouri action in the chiral limit takes the intriguing form
\begin{align}
S^{\beta=-i}_{\text{H-MM}}(\mathcal{A})=\frac{i}{\kappa}\int_{M}{\braket{F(\mathcal{A^+})\wedge\mathcal{E}}+\frac{1}{4L^2}\braket{\mathcal{E}\wedge\mathcal{E}}}+S_{\text{bdy}}(\mathcal{A}^+)
\label{D.eq:7.8}
\end{align}
with $\mathcal{E}$ the super electric field canonically conjugate to the super Asthekar connection $\mathcal{A}^+$ and transforming under the Adjoint representation of $\mathrm{OSp}(\mathcal{N}|2)_{\mathbb{C}}$. The boundary action $S_{\text{bdy}}(\mathcal{A}^+)$ of the theory is given by  
\begin{equation}
S_{\text{bdy}}(\mathcal{A}^+)\equiv S_{\mathrm{CS}}(\mathcal{A}^+)=\frac{k}{4\pi}\int_{H}{\braket{\mathcal{A}^+\wedge\mathrm{d}\mathcal{A}^++\frac{1}{3}\mathcal{A}^+\wedge[\mathcal{A}^+\wedge\mathcal{A}^+]}}
\label{D.eq:7.12}
\end{equation}
with $H:=\partial M$ and thus, in particular, corresponds to the action of a $\mathrm{OSp}(\mathcal{N}|2)_{\mathbb{C}}$ super Chern-Simons theory with (complex) Chern-Simons level $k=i4\pi L^2/\kappa=-i12\pi/\kappa\Lambda_{\text{cos}}$. As discussed in detail in \cite{Eder:2022,Eder:2021rgt}, this boundary action arising from \eqref{D.eq:6.4} in the chiral limit is indeed unique if one imposes supersymmetry invariance at the boundary (see also \cite{Andrianopoli:2014aqa,Andrianopoli:2020zbl}).

The decomposition of \eqref{D.eq:7.8} into a bulk and boundary action leads to an additional boundary condition coupling bulk and boundary degrees of freedom in order to ensure consistency with the equations of motion of the full theory. This boundary condition is given by
\begin{equation}
    \underset{\raisebox{1pt}{$\Longleftarrow$}}{F(\mathcal{A}^{+})}=-\frac{1}{2L^2}\underset{\raisebox{1pt}{$\Leftarrow$}}{\mathcal{E}}
    \label{D.eq:4.11.2}
\end{equation}
where the arrow denotes the pullback of the respective fields to the boundary. This condition will play a prominent role in the construction of the quantum theory of the full theory to be discussed in section \ref{OutlookBHentropy}.\\

Let us finally discuss some central aspects of the canonical description of the theory. The graded symplectic phase space of the canonical theory is generated by the canonically conjugate variables $(\mathcal{A}^{+ \underline{A}}_a,\mathcal{E}^a_{\underline{B}})$ with $\mathcal{A}^{\underline{A}}_a$ the coefficients of the super Asthekar connection w.r.t. a homogeneous basis $(T_{\underline{A}})_{\underline{A}}$ of $\mathfrak{osp}(\mathcal{N}|2)_{\mathbb{C}}$ and pulled back to the three-dimensional Cauchy slices $\Sigma$ of the globally hyperbolic spacetime manifold $M=\mathbb{R}\times\Sigma$. The canonically conjugate momentum $\mathcal{E}^a_{\underline{A}}$ is defined in terms of the super electric field $\mathcal{E}$ via
\begin{equation}
\mathcal{E}^a_{\underline{A}}:=\frac{1}{2}\epsilon^{abc}\mathscr{S}_{\underline{B}\underline{A}}\mathcal{E}^{\underline{B}}_{bc} 
\label{DH.eq:1.1}
\end{equation}
with $\mathscr{S}_{\underline{A}\underline{B}}:=\braket{T_{\underline{A}},T_{\underline{B}}}$. The pre-symplectic structure of the full theory including bulk and boundary degrees of freedom takes the form 
\begin{align}
    \Omega_{\Sigma}(\delta_1,\delta_2)&=\frac{2i}{\kappa}\int_{\Sigma}\braket{\delta_{[1}\mathcal{A}^+\wedge\delta_{2]}\mathcal{E}}-\frac{k}{2\pi}\int_{\Delta}\braket{\delta_{[1}\mathcal{A}^+\wedge\delta_{2]}\mathcal{A}^+}
    \label{D.eq:4.19.1}
\end{align}
From \eqref{D.eq:4.19.1}, it follows that the canonically conjugate variables indeed satisfy the graded Poisson relations
\begin{equation}
    \{\mathcal{E}^a_{\underline{A}}(x),\mathcal{A}_b^{\underline{B}}(y)\}=i\kappa\delta^a_b\delta^{\underline{B}}_{\underline{A}}\delta^{(3)}(x,y)
    \label{Chapter5:gradedPoisson}
\end{equation}
$\forall x,y\in\Sigma$. In case of a nontrivial boundary, in order to account for functional differentiability, it follows immediately from \eqref{D.eq:7.8} that the Gauss constraint is given by 
\begin{equation}
    \mathscr{G}[\alpha]=-\frac{i}{\kappa}\int_{\Sigma}\braket{\mathcal{E}\wedge D^{(\mathcal{A}^+)}\alpha}+\frac{i}{\kappa}\int_{\Delta}\braket{\mathcal{E},\alpha}
    \label{GaussN=1+bdy}
\end{equation}
with $\alpha$ some arbitrary smooth $\mathfrak{osp}(\mathcal{N}|2)_{\mathbb{C}}$-valued smearing function defined on $\Sigma$ and $\Delta$ defined as $\Delta:=\Sigma\cap H$. Using \eqref{Chapter5:gradedPoisson}, one deduces that the Gauss constraint satisfies the graded Poisson relations $\{\mathscr{G}[\alpha],\mathscr{G}[\beta]\}=\mathscr{G}[[\alpha,\beta]]$ and therefore generates local $\mathrm{OSp}(\mathcal{N}|2)_{\mathbb{C}}$ gauge transformations on phase space.\\
The boundary condition implies that the pre-symplectic structure $\Omega_{\Sigma}$ of the full theory is conserved, i.e., independent of the choice of a Cauchy hypersurface. To see this, let $\Sigma_i$ for $i=1,2$ be two Cauchy hypersurfaces and $B\subset H$ be a subset of the boundary enclosed by $\Sigma_1$ and $\Sigma_2$. Then, since \emph{on-shell} the pre-symplectic current of the bulk pre-symplectic structure defines a closed 2-form on field space \cite{Geiller:2017xad}, by Stokes' theorem, it follows that  
 \begin{align}
     \Omega_{\Sigma_2}(\delta_1,\delta_2)-\Omega_{\Sigma_1}(\delta_1,\delta_2)=&-\frac{2i}{\kappa}\int_{B}\braket{\delta_{[1}\mathcal{A}^+\wedge\delta_{2]}\mathcal{E}}-\frac{2iL^2}{\kappa}\int_{\Delta_2}\braket{\delta_{[1}\mathcal{A}^+\wedge\delta_{2]}\mathcal{A}^+}\nonumber\\
     &+\frac{2iL^2}{\kappa}\int_{\Delta_1}\braket{\delta_{[1}\mathcal{A}^+\wedge\delta_{2]}\mathcal{A}^+}
     \label{D.eq:PreSym1}
 \end{align}
with $\Delta_i:=\Sigma_i\cap H$ for $i=1,2$. According to boundary condition \eqref{D.eq:4.11.2}, the variation of the super electric field $\mathcal{E}$ on $B$ is given by $\delta\mathcal{E}|_B=-2L^2\delta F(\mathcal{A}^+)|_B=-2L^2D^{(\mathcal{A}^+)}\delta\mathcal{A}^+|_B$. Hence, this implies that first term on the right-hand side of Eq. \eqref{D.eq:PreSym1} can be written as
\begin{align}
-\frac{2i}{\kappa}\int_{B}\braket{\delta_{[1}\mathcal{A}^+\wedge\delta_{2]}\mathcal{E}}&=\frac{2iL^2}{\kappa}\int_{B}\mathrm{d}\!\braket{\delta_{[1}\mathcal{A}^+\wedge\delta_{2]}\mathcal{A}^+}\nonumber\\
&=\frac{2iL^2}{\kappa}\int_{\Delta_2}\braket{\delta_{[1}\mathcal{A}^+\wedge\delta_{2]}\mathcal{A}^+}-\frac{2iL^2}{\kappa}\int_{\Delta_1}\braket{\delta_{[1}\mathcal{A}^+\wedge\delta_{2]}\mathcal{A}^+}
\label{D.eq:PreSym2}
 \end{align}
Thus, when inserted back into \eqref{D.eq:PreSym1}, it follows immediately that the individual terms on the right-hand side cancel exactly finally proving that, on shell, $\Omega_{\Sigma_2}(\delta_1,\delta_2)=\Omega_{\Sigma_1}(\delta_1,\delta_2)$, that is, the pre-symplectic structure of the full theory is indeed conserved.

\section{Quantum theory}
According to the discussion in the previous chapter, in the chiral limit, the phase space of AdS Holst-supergravity turns out to be a graded generalization of the purely bosonic theory. Hence, this suggests to canonically quantize the theory adapting and generalizing tools from standard LQG. In the following sections, let us illustrate the construction of the so-called graded holomomy-flux algebra as well as the quantum theory corresponding to a representation of this superalgebra on a super Hilbert space. For more details including a mathematically consistent analysis using the concept of enriched categories which takes into account the proper implementation of the anticommutative nature of fermionic fields, we refer to \cite{Eder:2022}.  

\subsection{The graded holonomy-flux algebra}\label{sec:qt}
The super Ashtekar connection defines a super connection 1-form on an associated $\mathcal{G}:=\mathrm{OSp}(\mathcal{N}|2)_{\mathbb{C}}$-bundle. Hence, it follows that $\mathcal{A}^+$ induces holonomies, i.e., parallel transport maps\footnote{In fact, in order to consistently incorporate the antiommutative nature of fermionic field, it turns out that one actually has to work in an enriched category of supermanifolds. As a result, it follows that the fields are parametrized by an additional parametrizing supermanifold $\mathcal{S}$. Hence, as a consequence, this implies that holonomies have to be interpretated as $\mathcal{S}$-points, i.e., morphisms $h_e[\mathcal{A}^+]:\,\mathcal{S}\rightarrow\mathcal{G}$ which in turn can be regarded as group elements of a generalized super Lie group $\mathcal{G}(\mathcal{S})$ (see \cite{Eder:2022} for more details).} $h_e[\mathcal{A}^+]\in\mathcal{G}$ along one-dimensional paths $e$ embedded in $\Sigma$. For any two composable smooth paths $e,e'$ embedded in $\Sigma$, the holonomy satisfies 
\begin{equation}
   h_{e\circ e'}[\mathcal{A}^+]=h_{e'}[\mathcal{A}^+]\circ h_e[\mathcal{A}^+] 
\end{equation}
Hence, the holonomy induces a contravariant functor
\begin{equation}
    H:\,\mathbf{P}(\Sigma)\rightarrow\mathbf{G},\,e\mapsto h_e[\mathcal{A}^+]
\end{equation}
from the path groupoid $\mathbf{P}(\Sigma)$ to the gauge groupoid $\mathbf{G}$ with points in $\Sigma$ as objects and arrows $x\rightarrow y$ between points $x,y\in\Sigma$ labeled by group elements $g\in\mathcal{G}$.\\
As common in LQG, for the construction of the classical algebra, in the following we consider the whole set\footnote{Here, $\mathbf{Cat}$ denotes the \emph{category of small categories} with small categories $\mathcal{C}$ as objects and covariant functors $F:\,\mathcal{C}\rightarrow\mathcal{D}$ between small categories as morphisms where a category $\mathcal{C}$ is called \emph{small} if the collection of objects $\mathbf{Ob}(\mathcal{C})$ defines a set. This category can be even lifted to a \emph{2-category} regarding natural transformations $\eta:\,F\rightarrow G$ between functors as \emph{2-morphisms}.} $\mathrm{Hom}_{\mathbf{Cat}}(\mathbf{P}(\Sigma)^{\mathrm{op}},\mathbf{G})$, that is, the set of \emph{all} contravariant functors $H:\,\mathbf{P}(\Sigma)\rightarrow\mathbf{G}$ from the path groupoid to the gauge groupoid $\mathbf{G}$. That is, we do not restrict to those functors arising from the parallel transport map of a smooth super connection 1-form. For this reason, we will also refer to a such functor $H$ as a \emph{generalized super connection}. Next, we are looking for a different description of the set of generalized super connections on the whole path goupoid $\mathbf{P}(\Sigma)$ in terms of subsets defined on subgroupoids $l(\gamma)$ generated by finite \emph{graphs} $\gamma$. The collection $\mathcal{L}$ of all such subgroupoids $l$ forms a partially ordered set $\mathcal{L}\equiv(\mathcal{L},\leq)$ where $l\leq l'$ for any $l,l'\in\mathcal{L}$ iff $l$ is a subgroupoid of $l'$. In the following, let us assume that $\mathcal{L}$ is directed, i.e., $\forall l,l'\in\mathcal{L}$, there exists $l''\in\mathcal{L}$ such that $l,l'\leq l''$. For this to be true, similarly as in \cite{Thiemann:2007pyv,Lewandowski:2005jk}, one probably needs to work in a semi-analytic category of supermanifolds (see Remark 5.5.1 in \cite{Eder:2022}).\\
As in the purely bosonic theory, it follows that contravariant functors $H:\,\mathbf{P}(\Sigma)\rightarrow\mathbf{G}$ defined on the whole path groupoid $\mathbf{P}(\Sigma)$ can equivalently be described in terms of their restrictions $H|_l$ on subgroupoids $l\in\mathcal{L}$. As explained in detail in \cite{Eder:2022}, this also enables one to equip this set with a topology which, under certain assumptions on the gauge group $\mathcal{G}$, turns out to be \emph{projectively Hausdorff}. For this, for any $l\in\mathcal{L}$, we define
\begin{equation}
    \EuScript{A}_{l}:=\mathrm{Hom}_{\mathbf{Cat}}(l^{\mathrm{op}},\mathbf{G})
    \label{DH.eq:1.8}
\end{equation}
It is clear that a contravariant functor $H$ on a subgroupoid $l\equiv l(\gamma)$ generated by a graph $\gamma$ is uniquely determined by its images $(H(e_i))_{i=1,\ldots,n}$ of the underlying edges $e_i$. Hence, this yields a bijection
\begin{equation}
    \EuScript{A}_{l}\stackrel{\sim}{\rightarrow}\mathcal{G}^{|E(\gamma)|},\,H\mapsto (H(e_1),\ldots,H(e_n))
    \label{DH.eq:1.9}
\end{equation}
For any $l,l'\in\mathcal{L}$ with $l\leq l'$, one has a surjective mapping
\begin{equation}
    p_{ll'}:\,\EuScript{A}_{l'}\rightarrow\EuScript{A}_{l}
    \label{DH.eq:1.12}
\end{equation}
by simply restricting functors defined on $l'$ to the subgroupid $l$. In this way, one obtains a \emph{projective family} $(\EuScript{A}_{l},p_{ll'})_{l,l'\in\mathcal{L}}$ to which one can associate the corresponding \emph{projective limit}
\begin{equation}
    \overline{\EuScript{A}}:=\lim_{\longleftarrow}\,\EuScript{A}_{l}:=\{(H_l)_{l\in\mathcal{L}}\in\prod_{l\in\mathcal{L}}\EuScript{A}_{\mathcal{S},l}|\,p_{ll'}(H_{l'})=H_l\,\forall l\leq l'\}
    \label{DH.eq:1.13}
\end{equation}
which, as explained in \cite{Eder:2022}, intriguingly carries the structure of a Molotko-Sachse-type supermanifold. One can then prove that, via restriction of functors, this in fact yields a bijection
\begin{equation}
    \mathrm{Hom}_{\mathbf{Cat}}(\mathbf{P}(\Sigma)^{\mathrm{op}},\mathbf{G})\stackrel{\sim}{\rightarrow}\overline{\EuScript{A}},\,H\rightarrow(H|_l)_{l\in\mathcal{L}}
    \label{DH.eq:1.14}
\end{equation}
Using the identification \eqref{DH.eq:1.9}, for any $l\equiv l(\gamma)\in\mathcal{L}$, let us introduce a set of smooth functions on $\EuScript{A}_{l}$ denoted by $\mathrm{Cyl}^{\infty}(\EuScript{A}_{l})$ such that
\begin{equation}
\mathrm{Cyl}^{\infty}(\EuScript{A}_{l})\cong H^{\infty}(\mathcal{G}^{|E(\gamma)|},\mathbb{C})\cong H^{\infty}(\mathcal{G},\mathbb{C})^{\hat{\otimes}_{\pi}|E(\gamma)|}  
\label{DH.eq:1.17}
\end{equation}
where $H^{\infty}(\mathcal{G},\mathbb{C}):=H^{\infty}(\mathcal{G})\otimes\mathbb{C}$ is the super vector space of supersmooth functions on $\mathcal{G}$. Then, for any $l,l'\in\mathcal{L}$ with $l\leq l'$, the pullback of the projection \eqref{DH.eq:1.12} induces a map $p_{ll'}^*:\,\mathrm{Cyl}^{\infty}(\EuScript{A}_{l})\rightarrow\mathrm{Cyl}^{\infty}(\EuScript{A}_{l'})$. Thus, this in turn induces an \emph{inductive family} $(\mathrm{Cyl}^{\infty}(\EuScript{A}_{l}),p_{ll'}^*)_{l,l'\in\mathcal{L}}$ to which we can associate the corresponding \emph{inductive limit}
\begin{equation}
    \mathrm{Cyl}^{\infty}(\overline{\EuScript{A}}):=\lim_{\longrightarrow}\,\mathrm{Cyl}^{\infty}(\EuScript{A}_{l}):=\coprod_{l\in\mathcal{L}}\mathrm{Cyl}^{\infty}(\EuScript{A}_{l})_{\big{/}\sim}
    \label{DH.eq:1.18}
\end{equation}
which we will call the space of \emph{cylindrical functions} on $\overline{\EuScript{A}}$. In \eqref{DH.eq:1.18}, for two functions $f_l\in\mathrm{Cyl}^{\infty}(\EuScript{A}_{l})$ and $f_{l'}\in\mathrm{Cyl}^{\infty}(\EuScript{A}_{l'})$, the equivalence relation is defined via $f_l\sim f_{l'}$ iff there exists $l,l'\leq l''$ such that $p_{ll''}^*f_l=p_{l'l''}^*f_{l'}$.\\
\\
Next, let us turn to the dual dynamical variables given by the super electric field $\mathcal{E}$. Since it defines a 2-form, one can smear it over two dimensional surfaces embedded in $\Sigma$. Hence, let $S\subset\Sigma$ be a two-dimensional orientable submanifold which, in addition, we assume to be semianalytic and $n:\,S\rightarrow\mathfrak{g}$ be a $\mathfrak{g}$-valued smearing function defined on $S$. Then, we can integrate the super electric field over $S$ yielding the Grassmann-valued quantity
\begin{equation}
    \mathcal{E}_n(S):=\int_{S}{\braket{n,\mathcal{E}}}
    \label{DH.eq:1.19}
\end{equation}
which w.r.t. a local coordinate neighborhood $\phi:\,\mathbb{R}^3\supset U\rightarrow\phi(U)\subset\Sigma$ of $\Sigma$ adapted to $S$ such that, for sake of simplicity, $S\subset \phi(U)$, explicitly takes the form
\begin{equation}
    \mathcal{E}_n(S)=\int_{U}{\phi^*\braket{n,\mathcal{E}}}=\int_U{\frac{1}{2}n^{\underline{A}}\mathscr{S}_{\underline{B}\underline{A}}\mathcal{E}^{\underline{B}}_{ab}\,\mathrm{d}\phi^a\wedge\mathrm{d}\phi^b}=\int_U{\mathrm{d}^2u\,\frac{1}{2}n^{\underline{A}}\mathcal{E}^{c}_{\underline{A}}\epsilon_{cab}\partial_{u^1}\phi^a\partial_{u^2}\phi^b}
    \label{DH.eq:1.20}
\end{equation}
Via the graded Poisson bracket, it follows that the smeared quantities $\mathcal{E}_n(S)$ induce derivations $\mathcal{X}(S):\,\mathrm{Cyl}^{\infty}(\overline{\EuScript{A}})\rightarrow\mathrm{Cyl}^{\infty}(\overline{\EuScript{A}})$ on the space of cylindrical functions which we will call \emph{super electric fluxes}. On super holonomies $h_e[\mathcal{A}^+]$, their action is given by 
\begin{equation}
    \mathcal{X}_n(S)(h_e[\mathcal{A}]):=\{\mathcal{E}_n(S),h_e[\mathcal{A}]\}
    \label{DH.eq:1.21}
\end{equation}
As demonstrated in \cite{Eder:2022}, from \eqref{DH.eq:1.21} it follows that the action of $\mathcal{X}_n(S)$ on cylindrical functions $f_{l}\in\mathrm{Cyl}^{\infty}(\overline{\EuScript{A}}_l)$ associated to a subgroupoid $l\equiv l(\gamma)$ generated by a graph $\gamma$ adapted to $S$, this yields
\begin{equation}
    \mathcal{X}_n(S)(f_{l})=\frac{i\kappa}{4}\sum_{e\in E(\gamma),\,e\cap S\neq\emptyset}\epsilon(e,S)n^{\underline{A}}(b(e))R^e_{\underline{A}}f_{l}
    \label{DH.eq:1.27}
\end{equation}
where we used the identification $\EuScript{A}_{l}\cong\mathcal{G}^{|E(\gamma)|}$ such that $R^e_{\underline{A}}$ denotes the right-invariant vector field generated by $T_{\underline{A}}$ acting on the copy of $\mathcal{G}$ labeled by $e$ \cite{Thiemann:2007pyv}. From identity \eqref{DH.eq:1.27}, one deduces the remarkable property that, for a given graph $\gamma$ in $\Sigma$ generating the subgroupoid $l\equiv l(\gamma)$, super electric fluxes corresponding to surfaces $S$ which intersect the underlying edges only at their endpoints leave the space $\mathrm{Cyl}^{\infty}(\EuScript{A}_{l})$ of cylindrical functions on $\EuScript{A}_{l}$ invariant. Hence, if $V^{\infty}(\EuScript{A}_{l})$ denotes the superalgebra generated by the graded commutator of all such super electric flux operators, on this graph, we can define the \emph{graded holonomy-flux algebra} $\mathfrak{A}^{\mathrm{gHF}}_{l}$ via

\begin{equation}
    \mathfrak{A}^{\mathrm{gHF}}_{l}:=\mathrm{Cyl}^{\infty}(\EuScript{A}_{l})\rtimes V^{\infty}(\EuScript{A}_{l})
    \label{DH.eq:1.28}
\end{equation}
which, in particular, forms a (infinite-dimensional) super Lie algebra according to
\begin{equation}
    [(f,X),(f',Y)]:=(X(f')-(-1)^{|Y||f|}Y(f),[X,Y])
    \label{DH.eq:1.29}
\end{equation}
for any $f,f'\in\mathrm{Cyl}^{\infty}(\EuScript{A}_{l})$ and fluxes $X,Y\in V^{\infty}(\EuScript{A}_{l})$. Here, the parity $|X|$ of a homogeneous super electric flux $X$ is defined in the usual way regarding it as a homogeneous derivation on $\mathrm{Cyl}^{\infty}(\EuScript{A}_{l})$. Thus, for instance, in case $X\equiv\mathcal{X}_n(S)$ with $\mathcal{X}_n(S)$ defined via \eqref{DH.eq:1.27}, one has $|X|=|n|$ with $|n|=:i\in\mathbb{Z}_2$ the parity of the homogeneous smearing function $n:\,S\rightarrow\mathfrak{g}_i$.\\

More generally, considering all possible graphs, we define the \emph{graded holonomy-flux algebra} $\mathfrak{A}^{\mathrm{gHF}}$ via
\begin{equation}
    \mathfrak{A}^{\mathrm{gHF}}:=\mathrm{Cyl}^{\infty}(\overline{\EuScript{A}})\rtimes V^{\infty}(\overline{\EuScript{A}})
    \label{DH.eq:1.30}
\end{equation}
with $V^{\infty}(\overline{\EuScript{A}})$ the superalgebra generated by the graded commutator of super electric fluxes on the inductive limit $\mathrm{Cyl}^{\infty}(\overline{\EuScript{A}})$. Again, it follows that \eqref{DH.eq:1.30} forms a super Lie algebra. In context of the non-supersymmetric theory, this algebra is usually considered for quantization.\\

So far, we have not imposed any *-relation on the superalgebras \eqref{DH.eq:1.30} resp. \eqref{DH.eq:1.29} so that they form *-algebras. This is, however, necessary in order to identify physical quantities in terms of self-adjoint elements. In the context of chiral supergravity, it follows that the super Asthekar connection and its canonical conjugate momentum $\mathcal{E}$ have to satisfy certain reality conditions in order to ensure consistency with the equations of motions of oridinary \emph{real} supergravity. By re-expressing the reality conditions in terms of holonomy and flux variables, these may be used in order to impose *-relations the graded holonomy-flux algebra. But, since they are highly non-linear, even in the purely bosonic theory, this turns out to be a nontrivial task. Hence, in the following, we do not want to comment further on the specific form of the reality conditions and the *-relations imposed on the graded holonomy-flux algebra. Nevertheless, let us note that, in the context of a symmetry reduced model, we have been able to find an explicit form of the *-relation and to implement them rigorously in the quantum theory (see \cite{Eder:2020okh}).

\subsection{The bulk super Hilbert space of chiral LQG}
Having derived the graded generalization of the well-known holonomy-flux algebra in LQG, we would like to discuss the quantization of the theory studying representation of this algebra on a super Hilbert space (see also \cite{Eder:2022} and references therein for more details as well as a proper definition of super Hilbert spaces).\\
However, there, one runs into several problems as the underlying gauge supergroups given by the (complex) orthosymplectic supergroups $\mathrm{OSp}(\mathcal{N}|2)_{\mathbb{C}}$ are non-compact. Moreover, one also needs to deal with the consistent implementation of the reality conditions as one is still dealing with a complex theory. An interesting and elegant possibility to solve the reality conditions would be to be adapt the ideas of \cite{Thiemann:1995ug} and to introduce some kind of a Wick rotation on the phase space so that the complex theory arises from an Euclidean counterpart corresponding to a real Barbero-Immirzi parameter $\beta\in\{\pm 1\}$ via a Wick transformation. But, the resulting gauge group given by the real orthosymplectic supergroup $\mathrm{OSp}(\mathcal{N}|2)$ is still non-compact.\\  
Adapting ideas in context of the purely bosonic theory (see for instance \cite{Frodden:2012dq,Han:2014xna,Achour:2014eqa,Achour:2015zmk,Bodendorfer:2013hla} and references therein for recent advances in this direction), this may be solved by going over instead to their corresponding compact form given by unitary orthosymplectic group
\begin{equation}
\mathrm{UOSp}(\mathcal{N}|2)=\mathrm{OSp}(\mathcal{N}|2)\cap\mathrm{U}(\mathcal{N}|2)    
\end{equation} 
As already mentioned in the previous section, for the special case $\mathcal{N}=1$, besides compactness, this group has very useful properties such as the existence of an invariant Haar measure with respect to which, in particular, the unit function is normalizable which is important in context of loop quantization in order to implement cylindrical consistency. Nevertheless, this last property turns out to be no longer satisfied in case of extended supersymmetry corresponding to higher $\mathcal{N}>1$.\\ 
Anyway, since, we want to explicitly include the extended case  $\mathcal{N}=2$, in what follows, we will not discuss the question of how to impose cylindrical consistency and instead work on a single graph $\gamma$ in $\Sigma$. As argued in \cite{Achour:2015zmk}, we may therefore assume that the graph under consideration is at least suitably fine enough to resolve the topology of $\Sigma$. Let $\mathfrak{A}_{\gamma}^{\mathrm{gHF}}:=\mathfrak{A}_{l(\gamma)}^{\mathrm{gHF}}$ denote the graded holonomy-flux algebra w.r.t. the graph $\gamma$ and underlying gauge group given by $\mathrm{OSp}(\mathcal{N}|2)_{\mathbb{C}}$. The quantization of the theory then corresponds to a representation
\begin{equation}
    \pi_{\gamma}:\,\mathfrak{A}_{\gamma}^{\mathrm{gHF}}\rightarrow\mathrm{Op}(\mathcal{D}_{\gamma},\mathfrak{H}_{\gamma}^{\mathrm{cLQSG}})
    \label{DH.eq:3.10}
\end{equation}
of $\mathfrak{A}_{\gamma}^{\mathrm{gHF}}$ on the space of (un)bounded operators on a super Hilbert space $\mathfrak{H}_{\gamma}^{\mathrm{cLQSG}}$ mutually defined on a dense graded subspace $\mathcal{D}_{\gamma}\subset\mathfrak{H}_{\gamma}^{\mathrm{cLQSG}}$. To construct this representation, as pre-Hilbert space, we consider the super vector space $V_{\gamma}:=\mathrm{Cyl}^{\infty}(\EuScript{A}_{l(\gamma)})$ which, according to \eqref{DH.eq:1.17}, can be identified with
\begin{equation}
   H^{\infty}(\mathcal{G}^{|E(\gamma)|},\mathbb{C})\cong H^{\infty}(\mathcal{G},\mathbb{C})^{\otimes_{\pi} |E(\gamma)|}
    \label{DH.eq:3.11}
\end{equation}
or a suitable subspace thereof, if one restricts, for instance, to holomorphic functions as naturally arising from super holonomies induced by the super Ashtekar connection (see discussion below). Following the standard procedure in the purely bosonic theory, for the quantization, we choose a \emph{Ashtekar-Lewandowski-type representation} of $\mathfrak{A}^{\mathrm{gHF}}$ by setting
\begin{equation}
    \pi_{\gamma}(f_{\gamma}):=\widehat{f}_{\gamma},\quad\pi_{\gamma}(\mathcal{X}_n(S)):=i\hbar\mathcal{X}_n(S)
    \label{DH.eq:3.2}
\end{equation}
where $\widehat{f}_{\gamma}$ acts as a multiplication operator by $f_{\gamma}$.

For the super scalar product $\mathscr{S}$ on $V_{\gamma}$ we make the ansatz
\begin{equation}
\mathscr{S}(f|g):=\int_{\mathrm{SL}(2,\mathbb{C})}\mathrm{d}\mu_{\mathrm{SL}(2,\mathbb{C})}(g,\bar{g})\int_B\mathrm{d}\theta^A\mathrm{d}\bar{\theta}^{A'}\rho(g,\bar{g},\theta,\bar{\theta})\,\bar{f}g
\label{DH.eq:3.12}
\end{equation}
with $\mathrm{d}\mu_{\mathrm{SL}(2,\mathbb{C})}$ the invariant Haar measure on the underlying bosonic Lie group $\mathrm{SL}(2,\mathbb{C})$ and $\int_B$ the Berezin integral. Here, $\rho\equiv\rho(g,\bar{g},\theta,\bar{\theta})$ denotes an additional density which has been chosen in order to deal with the non-compactness of the group. In this context, note that, generically, the matrix coefficients of the super holonomies, as part of the underlying algebra and thus of the resulting state space in the quantum theory, are functions of the form
\begin{equation}
f=\sum_{\underline{I}}f_{\underline{I}}\psi^{\underline{I}}=f_{\emptyset}+f_A\psi^A+\frac{1}{2}f_{+-}\psi_A\psi^A
\label{DH.eq:3.13}
\end{equation}
with $f_{\underline{I}}$ Grassmann extensions of holomorphic functions on  $\mathrm{SL}(2,\mathbb{C})$. But, by Liouville's theorem, if required to be nontrivial, general functions of this kind cannot be of compact support. This is of course problematic in context of integration theory and thus for the proper definition of the inner product. Hence, either one excludes holomorphic functions already in the definition of the classical algebra or the measure on $\mathrm{SL}(2,\mathbb{C})$ is changed appropriately by introducing a density $\rho$ which is of compact support. The last possibility has been studied in \cite{Eder:2020okh} in the context of symmetry reduced models. There, the measure turns out to be in fact distributional. In particular, it was shown that this also enables one to exactly implement the reality conditions in the quantum theory. In context of the full theory with ordinary self-dual variables, this idea also been studied in \cite{Robert:2021} considering a specific subclass of the full reality conditions where it was found that the resulting density imposes a gauge-fixing onto the compact subgroup $\mathrm{SU}(2)$ of $\mathrm{SL}(2,\mathbb{C})$. Maybe, these results can be extended to the supersymmetric setting possibly involving the unitary orthosymplectic group $\mathrm{UOSp}(1|2)$ which, as explained above, has many interesting properties quite analogous to the purely bosonic theory. In fact, this group will play an important role in the context of the entropy computation to be discussed in Section \ref{sec:EntropyBH}\\
\\
Ultimately, for the construction of the super Hilbert space, we have to choose an endomorphism $J:\,V_{\gamma}\rightarrow V_{\gamma}$ such that the induced inner product $\braket{\cdot|\cdot}_{J}:=\mathscr{S}(\cdot|J\cdot)$ is positive definite. The choice of such an endomorphism is, of course, not unique but strongly restricted by the correct implementation of the reality conditions (see \cite{Eder:2022} as well as \cite{Eder:2020okh} in the context of symmetry reduced models). Using this inner product, we can then complete $V_{\gamma}$ to a Hilbert space $\mathfrak{H}_{\gamma}^{\mathrm{cLQSG}}$ so that finally end up with the super Hilbert space $(\mathfrak{H}_{\gamma}^{\mathrm{cLQSG}},\mathscr{S},J)$. 

\subsection{Super spin networks and the super area operator}\label{Section:chSUGRA-superspin}

Having constructed the Hilbert space representation of the classical algebra underlying canonical chiral supergravity, we next have to select the proper subspace of \emph{physical states} consisting of states in $\mathfrak{H}_{\mathcal{S},\gamma}^{\mathrm{cLQSG}}$ that are annihilated by the operators corresponding to the constraints of the canonical classical theory (see \cite{Eder:2022} for more details). In the following, let us only focus on the super Gauss constraint. In fact, the particular advantage of the loop representation as studied in this section is the rather straightforward implementation of the super Gauss constraint \eqref{GaussN=1+bdy} in the quantum theory implying invariance of physical states under local gauge transformations. 

To this end, note that the super Gauss constraint in the bulk theory can equally be written in the form
\begin{equation}
    \mathscr{G}[\alpha]=-\frac{i}{\kappa}\int_{\Sigma}\braket{D^{(\mathcal{A}^+)}\alpha\wedge\mathcal{\mathcal{E}}}=-\frac{i}{\kappa}\int_{\Sigma}\mathrm{d}^3x\,(D^{(\mathcal{A}^+)}_a\alpha^{\underline{A}})\mathcal{E}^a_{\underline{A}}=:-\frac{i}{\kappa}\mathcal{E}(D^{(\mathcal{A}^+)}\alpha)
    \label{DH.eq:3.14}
\end{equation}
and thus resembles the definition of a super electric flux but smeared over a three-dimensional region instead of two-dimensional surfaces. Thus, for the corresponding operator in the quantum theory, we may set
\begin{equation}
    \widehat{\mathscr{G}}[\alpha]:=\frac{\hbar}{\kappa}\{\mathcal{E}(D^{(\mathcal{A}^+)}\alpha),\cdot\}
    \label{DH.eq:3.15}
\end{equation}
Following the same steps as in the purely bosonic theory, it is then immediate to see that the super Gauss constraint operator takes the form
\begin{equation}
    \widehat{\mathscr{G}}[\alpha]=\frac{i\hbar}{2}\sum_{v\in V(\gamma)}\alpha^{\underline{A}}(v)\left[\sum_{e\in E(\gamma),b(e)=v}R_{\underline{A}}^e-\sum_{e\in E(\gamma),f(e)=v}L_{\underline{A}}^e\right]
    \label{DH.eq:3.16}
\end{equation}
In particular, due to its structure, the super Gauss constraint has a well-defined action on the super Hilbert space as it takes the standard form of a super electric flux operator and maps cylindrical functions to cylindrical functions. For a generic state $f\in\mathfrak{H}_{\mathcal{S},\gamma}^{\mathrm{cLQSG}}$ to be \emph{physical}, this then yields the condition
\begin{equation}
    \widehat{\mathscr{G}}[\alpha]f=0
    \label{DH.eq:3.17}
\end{equation}
that is, according to \eqref{DH.eq:3.16}, physical states have to be invariant under both the left- and right-regular representation of $\mathrm{OSp}(\mathcal{N}|2)$. 

In standard loop quantum gravity, one considers a typical class of states satisfying the constraint equation \eqref{DH.eq:3.17} given by the so-called \emph{spin network states}. These states are constructed via contraction of matrix coefficients of irreducible representations of the underlying gauge group. In fact, in case that the bosonic group is compact, it follows that these type of states form an orthonormal basis of the entire Hilbert space. This follows from the well-known Peter-Weyl theorem which is valid for compact bosonic groups. However, in case of general super Lie groups such a general statement, unfortunately, is not known.

For the construction of the spin network states, it is crucial that the representations under consideration form a tensor category. We may call such kind of representations having this property \emph{admissible} in what follows. Thus, in the supersymmetric setting, by restricting to admissible representations of the underlying gauge supergroup, one is able to construct invariant states in the theory. This leads to the notion of \emph{super spin network states}. For $\mathcal{N}=1$ and considering finite-dimensional representations, these have been studied for instance in the References \cite{Gambini:1995db,Ling:1999gn}. In fact, the finite-dimensional irreducible representations of the orthosymplectic series $\mathrm{OSp}(\mathcal{N}|2)$ for $\mathcal{N}=1,2$ are very well-known and have been intensively studied in the literature (see e.g. \cite{Scheunert:1976wj,Scheunert:1976wi,Minnaert:1990sz,Berezin:1981} as well as Section \ref{section 4.2} below ). In particular, for the case $\mathcal{N}=1$, it follows that these type of representations form a subcategory which is closed under tensor product. In fact, the same applies to the extended case $\mathcal{N}=2$ if one restricts to a particular subclass of the so-called \emph{typical representations} (see \cite{Scheunert:1976wj} for more details). For the rest of this section, we want to describe the construction of the super spin network states for such a suitable subclass of representations explicitly including the possibility of infinite-dimensional representations as well as the extended supersymmetric case $\mathcal{N}=2$. 

To this end, let $\EuScript{P}_{\mathrm{adm}}$ denote the set of equivalence classes of admissible irreducible representations (finite- or infinite-dimensional) of $\mathrm{OSp}(\mathcal{N}|2)$ with $\mathcal{N}=1,2$. For any subset $\Vec{\pi}:=\{\pi_{e}\}_{e\in E(\gamma)}\subset\EuScript{P}_{\mathrm{adm}}$, we then define the cylindrical function $T_{\gamma,\Vec{\pi},\Vec{m},\Vec{n}}\in\mathrm{Cyl}^{\infty}(\EuScript{A}_{\mathcal{S},\gamma})$ via
\begin{equation}
    T_{\gamma,\Vec{\pi},\Vec{m},\Vec{n}}:=\prod_{e\in E(\gamma)}\tensor{(\pi_e)}{^{m_e}_{n_e}}
    \label{DH.eq:3.18}
\end{equation}
also called a \emph{gauge-variant super spin network state} where, for any edge $e\in E(\gamma)$, $\tensor{(\pi_e)}{^{m_e}_{n_e}}$ denote certain matrix coefficients of the representation $\pi_e\in\EuScript{P}_{\mathrm{adm}}$. By definition, it then follows from the general transformation law of a super holonomy under local gauge transformations (see \cite{Eder:2022}), that, at each vertex $v\in V(\gamma)$, the state \eqref{DH.eq:3.18} transforms under the following tensor product representation of $\mathrm{OSp}(\mathcal{N}|2)$
\begin{equation}
    \pi'_v:=\left(\bigotimes_{e\in I(v)}\pi_e\right)\otimes\left(\bigotimes_{e\in F(v)}\pi_{e}^*\right)
    \label{DH.eq:3.19}
\end{equation}
where $\pi^*_e\in\EuScript{P}_{\mathrm{adm}}$ denotes the right dual representation corresponding to $\pi_e$. Here, $I(v)$ and $F(v)$ are defined as subsets of $E(\gamma)$ consisting  of all edges $e\in E(\gamma)$ which are beginning or ending at the vertex $v\in V(\gamma)$, respectively. Hence, in order to construct gauge-invariant states, at each vertex $v\in V(\gamma)$, we have to assume that the trivial representation $\pi_0$ appears in the decomposition of the product representation \eqref{DH.eq:3.19}, i.e., $\pi_0\in\pi'_v$ $\forall v\in V(\gamma)$. For any $v\in V(\gamma)$, we can then choose an intertwiner $I_v$ which contracted with the state \eqref{DH.eq:3.19} project onto the trivial representation at any vertex. As a consequence, the resulting state transforms trivially under local gauge transformations and thus indeed forms a gauge-invariant state which we call a \emph{(gauge-invariant) super spin network state}.
 
On the super Hilbert space $\mathfrak{H}^{\mathrm{cLQSG}}_{\mathcal{S},\gamma}$, one can introduce a gauge-invariant quantity in analogy to the area operator in ordinary LQG. More precisely, since the super electric field $\mathcal{E}$ defines a $\mathrm{Lie}(\mathcal{G})$-valued 2-form, for any oriented (semianalytic) surface $S$ embedded in $\Sigma$, one can define the \emph{graded} or \emph{super area} $\mathrm{gAr}(S)$ via
\begin{equation}
\mathrm{gAr}(S):=\alpha\int_{S}\|\mathcal{E}\|
\label{Chapter5:area1}
\end{equation}
with $\alpha\in\mathbb{R}_{+}$ an arbitrary positive real number. Here, generalizing the considerations in \cite{Eder:2018uzm,Corichi:2000dm,Ashtekar:2000hw} in the context of the purely bosonic theory to the supersymmetric setting, the norm $\|\mathcal{E}\|$ is a 2-form on $S$ defined as follows: Let $\iota_S:\,S\hookrightarrow\Sigma$ denote the embedding of the surface $S$ in $\Sigma$. Since, $\iota^*_S\mathcal{E}$ defines a 2-form on $S$, it follows that there exists a unique $\mathrm{Lie}(\mathcal{G})$-valued function $\mathcal{E}_S:\,\mathcal{S}\times S\rightarrow\mathrm{Lie}(\mathcal{G})$ such that $\iota^*_S\mathcal{E}=\mathcal{E}_S\,\mathrm{vol}_S$. The norm $\|\mathcal{E}\|$ is then given by
\begin{equation}
    \|\mathcal{E}\|:=\sqrt{\braket{\mathcal{E}_S,\mathcal{E}_S}}
    \label{Chapter5:area2}
\end{equation}
For  the special case $\mathcal{N}=1$, it follows that the expression \eqref{Chapter5:area1} coincides with the super area as considered in \cite{Ling:1999gn}. Note that, in case that the underlying parametrizing supermanifold is chosen to be trivial $\mathcal{S}=\{*\}$, i.e. the fermionic degrees of freedom vanish, the super area reduces to the standard area of $S$ in Riemannian geometry provided that for the constant $\alpha$ one sets $\alpha=\sqrt{2}$. 

By definition, the quantity \eqref{Chapter5:area1} solely depends on the super electric field which defines a phase space variable. Thus, we can implement it in the quantum theory. To do so, we first need to perform an appropriate regularization. Following \cite{Eder:2018uzm}, let us therefore assume that the surface $S$ intersects the graph $\gamma$ only in its vertices and is contained within a single coordinate neighborhood $(U,\phi_U)$ of $\Sigma$ adapted to $S$. Furthermore, let $\mathcal{U}_{\epsilon}=\{U_i\}_i$ be a partition of $U$ of fineness $\epsilon>0$ such that $S$ is covered by the $S_{U_i}:=\phi_{U}(U_i)$. Then, for $\epsilon>0$, we define
\begin{equation}
\mathrm{gAr}_{\epsilon}(S):=\sum_{V\in\mathcal{U}_{\epsilon}}{\|\mathcal{E}(S_V)\|}\equiv\sum_{V\in\mathcal{U}_{\epsilon}}{\sqrt{\mathscr{T}^{\underline{A}\underline{B}}\mathcal{X}_{\underline{B}}(S_V)\mathcal{X}_{\underline{A}}(S_V)}}
\label{Chapter5:area3}
\end{equation} 
where $\mathcal{X}_{\underline{A}}(S_V)$ denotes the super electric flux operator smeared over $S_V$ with smearing function $n:\,S\rightarrow\mathfrak{g}$ satisfying $n^{\underline{B}}\equiv 1$ for $\underline{B}=\underline{A}$ and $n^{\underline{B}}=0$ otherwise. In the limit $\epsilon\rightarrow 0$, this then implies $\mathrm{gAr}(S)=\lim_{\epsilon\rightarrow 0}\mathrm{gAr}_{\epsilon}(S)$. Using this regularization, we can define the \emph{super area operator} as follows 
\begin{equation}
\widehat{\mathrm{gAr}}(S)=\lim_{\epsilon\rightarrow 0}\widehat{\mathrm{gAr}}_{\epsilon}(S), \qquad 
\widehat{\mathrm{gAr}}_{\epsilon}(S)=\sum_{V\in\mathcal{U}_{\epsilon}}{\sqrt{\mathscr{T}^{\underline{A}\underline{B}}\widehat{\mathcal{X}}_{\underline{B}}(S_V)\widehat{\mathcal{X}}_{\underline{A}}(S_V)}}
\label{Chapter5:area4}
\end{equation}
Next, let us derive an explicit formula for its action on super spin network states. To this end, following again \cite{Eder:2018uzm} in the context of purely bosonic theory, we compute
\begin{align}
\mathscr{T}^{\underline{A}\underline{B}}\widehat{\mathcal{X}}_{\underline{B}}(S_V)\widehat{\mathcal{X}}_{\underline{A}}(S_V)&=\left(\frac{\hbar\kappa}{4}\right)^2\mathscr{T}^{\underline{A}\underline{B}}\left(\sum_{e\cap S_V\neq\emptyset}{\epsilon(e,S_V)R_{\underline{B}}^e}\right)\left(\sum_{e\cap S_V\neq\emptyset}{\epsilon(e,S_V)R_{\underline{A}}^e}\right)\nonumber\\
&=\left(\frac{\hbar\kappa}{4}\right)^2\mathscr{T}^{\underline{A}\underline{B}}\left(R^{\mathrm{in}}_{\underline{B}}-R^{\mathrm{out}}_{\underline{B}}\right)\left(R^{\mathrm{in}}_{\underline{A}}-R^{\mathrm{out}}_{\underline{A}}\right)\nonumber\\
&=\left(\frac{\hbar\kappa}{4}\right)^2\mathscr{T}^{\underline{A}\underline{B}}\left(2R^{\mathrm{in}}_{\underline{B}}R^{\mathrm{in}}_{\underline{A}}+2R^{\mathrm{out}}_{\underline{B}}R^{\mathrm{out}}_{\underline{A}}-\left(R^{\mathrm{in}}_{\underline{B}}+R^{\mathrm{out}}_{\underline{B}}\right)\left(R^{\mathrm{in}}_{\underline{A}}+R^{\mathrm{out}}_{\underline{A}}\right)\right)\nonumber\\
&=:-\left(\frac{\hbar\kappa}{4}\right)^2\left(2\Delta_{I}+2\Delta_F-\Delta_{I\cup F}\right)
\end{align}
with $R^{\mathrm{in}}_{\underline{A}}=\sum_{e\,\mathrm{ingoing}}{R_{\underline{A}}^e}$ and $R^{\mathrm{out}}_{\underline{A}}=\sum_{e\,\mathrm{outgoing}}{R_{\underline{A}}^e}$. Moreover, $\Delta:=-\mathscr{T}^{\underline{A}\underline{B}}R_{\underline{B}}R_{\underline{A}}$ denotes the \emph{super Laplace-Beltrami operator} of the super Lie group $\mathcal{G}$.\\
To simplify the expression, suppose that the surface $S$ intersects the graph $\gamma$ in a single divalent vertex $v\in V(\gamma)$ so that, at this vertex, one has $\Delta\equiv\Delta_I=\Delta_F$ as well as $\Delta_{I\cup F}=0$. If we identify $C_2^{\mathfrak{osp}}:=\frac{\Delta}{2}$ with the \emph{quadratic Casimir operator} of $\mathfrak{osp}(\mathcal{N}|2)$ (see Section \ref{section 4.2}), it then follows for $\alpha=\sqrt{2}$ that the super area operator takes the form
\begin{equation}
    \widehat{\mathrm{gAr}}(S)=-8\pi i\sqrt{C_2^{\mathfrak{osp}}}
    \label{graded area with Casimir}
\end{equation}
Using \eqref{graded area with Casimir}, let us compute the action of the super area operator on a (gauge-invariant) super spin network state $T_{\gamma,\Vec{\pi},\Vec{m},\Vec{n}}$ for the special case $\mathcal{N}=1$. In the case that the edges of the graph are labeled super spin quantum numbers $j\in\mathbb{C}$ corresponding to the principal series of $\mathrm{OSp}(1|2)$ as discussed in detail in Section \ref{section 4.2}, it follows from \eqref{eq:4.2.22} that the action of the super area operator is given by
\begin{equation}
\widehat{\mathrm{gAr}}(S) T_{\gamma,\Vec{\pi},\Vec{m},\Vec{n}}=-8\pi il_p^2\sqrt{j\left(j+\frac{1}{2}\right)}T_{\gamma,\Vec{\pi},\Vec{m},\Vec{n}}
\label{eq:3.3.49}
\end{equation}
with $j\in\mathbb{C}$ the spin quantum number labeling the edge $e\in E(\gamma)$ that intersects the vertex $v$. For $j\in\frac{N_0}{2}$, this coincides with the result of \cite{Ling:1999gn}.

\subsection{Boundary theory}\label{OutlookBHentropy}
So far, we have restricted to the quantization of the bulk degrees of freedom in the framework LQSG. As a next step, we would like to discuss the quantization of the full theory. To this end, let us first focus on the canonical description of the boundary theory.

As discussed in detail in \cite{Eder:2022}, the 2+1-split of the super Chern-Simons action takes the form
\begin{equation}
S_{\mathrm{CS}}(A)=\frac{k}{4\pi}\int_{\mathbb{R}}{\mathrm{d}t\int_{\Delta_t}{\braket{-\mathcal{A}\wedge\dot{\mathcal{A}}+2\mathcal{A}_0F(\mathcal{A})-\mathrm{d}(\mathcal{A}_0\mathcal{A})}}}
\label{D.eq:1.4.11}
\end{equation}
As a consequence, the pre-symplectic structure of the canonical theory is given by
\begin{align}
\Omega_{\mathrm{CS}}(\delta_1,\delta_2)=-\frac{k}{2\pi}\int_{\Delta}\braket{\delta_{[1}\mathcal{A}\wedge\delta_{2]}\mathcal{A}}
\label{D.eq:1.4.14}
\end{align}
for variations $\delta\mathcal{A}\in T\mathscr{A}_{\Delta}$ where $\mathscr{A}_{\Delta}$ denotes the space of smooth super connection 1-forms on the induced $\mathcal{G}$-principal bundle $\mathcal{E}:=\mathcal{P}|_{\Delta}$ over $\Delta$. Since the difference of two super connections defines an even horizontal 1-form of type $(\mathcal{G},\mathrm{Ad})$, it follows that $T_{\mathcal{A}}\mathscr{A}_{\Delta}$ at any $\mathcal{A}\in\mathscr{A}_{\Delta}$ can be identified with $T_{\mathcal{A}}\mathscr{A}_{\Delta}\cong\Omega^1(\Delta,\mathrm{Ad}(\mathcal{E}))_0$. For the graded Poisson bracket, one obtains 
\begin{equation}
\{\mathcal{A}^{\underline{A}}_{a}(x),\mathcal{A}^{\underline{B}}_{b}(y)\}=-\frac{2\pi}{k}\mathscr{S}^{\underline{A}\underline{B}}\epsilon_{ab}\delta^{(2)}(x,y)
\label{D.eq:1.4.15}
\end{equation}
where $\mathscr{S}^{\underline{A}\underline{B}}$ denotes the matrix components of the inverse super metric satisfying $\mathscr{S}_{\underline{C}\underline{A}}\mathscr{S}^{\underline{C}\underline{B}}=\delta_{\underline{A}}^{\underline{B}}$. Moreover, from the split action \eqref{D.eq:1.4.11}, we can read off the constraint
\begin{equation}
    \mathcal{F}[\alpha]:=\frac{k}{2\pi}\int_{\Delta}\braket{\alpha F(\mathcal{A})}
    \label{D.eq:1.4.16}
\end{equation}
which imposes the condition $F(\mathcal{A})=0$, that is, the curvature of the super connection on $\Delta$ is constrained to vanish. For this reason, $\mathcal{F}[\alpha]$ is also referred to as the \emph{flatness constraint}. Actually, since the curvature contains a term involving an exterior derivative, the flatness constraint \eqref{D.eq:1.4.16}, in general, turns out to be not functionally differentiable. In case that $\Delta$ has a nontrivial boundary $\partial\Delta$ which, in the context of two dimensions, we will refer to as the \emph{corner} of $\Delta$, one needs to require that the smearing function in \eqref{D.eq:1.4.16} satisfies the condition $\alpha|_{\partial\Delta}\equiv 0$.\\ 
In the framework of LQG, singularities on the boundary typically arise from the intersection of the boundary with spin network states. Assuming that the spin network edges piercing the boundary have some infinitesimal but \emph{nonzero} width, this induces infinitesimal holes at the punctures on the boundary, such that, at each puncture, $\partial\Delta$ becomes nontrivial and topologically equivalent to a 1-dimensional circle. As a consequence, this gives rise to new physical degrees of freedom on the boundary which are localised on the corner $\partial\Delta$. In the context of LQG, this was first observed in \cite{Ghosh:2014rra} and discussed more expansively, e.g., in \cite{Geiller:2017xad,Freidel:2016bxd,Freidel:2020xyx,Freidel:2020svx,Freidel:2020ayo}. As argued in \cite{Ghosh:2014rra}, based on a general proposal formulated in \cite{Carlip:2007qh,Carlip:2011vr}, these new degrees of freedom may also account for black hole entropy and thus may play a crucial role in the quantum description of the black holes. In fact, it turns out that these contain the physical degrees of freedom associated to the Hilbert spaces of conformal blocks which are usually considered in the context of black hole entropy computations in LQG.\\ 
\\
While we have not yet been able to complete the definition of the Hilbert space for chiral LQSG, extrapolating from what we have it seems that all these observations carry over quite naturally to the context of the quantum description of chiral supergravity with $\mathcal{N}$-extended supersymmetry. In that case, we have described in Section \ref{Section:chSUGRA-superspin} how the quantum excitations of the bulk degrees of freedom are represented by super spin network states associated to the gauge supergroup $\mathrm{OSp}(\mathcal{N}|2)_{\mathbb{C}}$. On the other hand, in Section \ref{sec:Review}, we have explained that the boundary theory is described in terms of a $\mathrm{OSp}(\mathcal{N}|2)_{\mathbb{C}}$ super Chern-Simons theory. Hence, it follows that, due to the quantization of super electric fluxes in the bulk, super spin network states induce singularities on the boundary. To see this, note that the Gauss constraint $\mathscr{G}_{\text{full}}[\alpha]$ of the full theory including both bulk and boundary degrees of freedom is given by the sum of the Gauss constraint \eqref{GaussN=1+bdy} in the bulk as well as the flatness constraint \eqref{D.eq:1.4.16} on the boundary, that is,
\begin{equation}
    \mathscr{G}_{\text{full}}[\alpha]=-\frac{i}{\kappa}\int_{\Sigma}\braket{D^{(\mathcal{A}^+)}\alpha\wedge\mathcal{E}}+\frac{i}{\kappa}\int_{\Delta}\braket{\alpha[\mathcal{E}-\frac{i\kappa k}{2\pi}F(\mathcal{A}^+)]}
    \label{Gauss:fullTheory}
\end{equation}
for any $\mathfrak{g}$-valued smearing function $\alpha$.\\
For a given finite graph $\gamma$ embedded in $\Sigma$, we define the Hilbert space $\mathfrak{H}_{\text{full},\gamma}$ w.r.t. $\gamma$ of the full theory as the tensor product
\begin{equation}
    \mathfrak{H}^{\text{full}}_{\gamma}=\mathfrak{H}^{\mathrm{cLQSG}}_{\gamma}\otimes\mathfrak{H}^{\text{CS}}_{\gamma}
    \label{Gauss:fullTheory1}
\end{equation}
with $\mathfrak{H}^{\mathrm{cLQSG}}_{\gamma}$ the Hilbert space of the quantized bulk degrees of freedom as constructed in Section \ref{Section:chSUGRA-superspin} and $\mathfrak{H}^{\text{CS}}_{\gamma}$ the Hilbert space corresponding to the quantized super Chern-Simons theory on the boundary.\\
As a next step, in order to implement the full Gauss constraint \eqref{Gauss:fullTheory} in the quantum theory, we have to regularize it over the graph $\gamma$. To this end, at each \emph{puncture} $p\in\EuScript{P}_{\gamma}:=\gamma\cap\Sigma$, let us choose a disk $D_{\epsilon}(p)$ on $\Delta$ around $p$ with radius $\epsilon>0$ and set 
\begin{equation}
    \mathcal{E}[\alpha](p):=\lim_{\epsilon\rightarrow 0}\int_{D_{\epsilon}(p)}\braket{\alpha,\mathcal{E}},\quad F[\alpha](p):=\lim_{\epsilon\rightarrow 0}\int_{D_{\epsilon}(p)}\braket{\alpha,F(\mathcal{A}^+)}
    \label{Gauss:fullTheory2}
\end{equation}
By definition, these quantities (or suitable functions thereof) can be promoted to well-defined operators in the quantum theory. Thus, it follows that the Gauss constraint operator of the full theory takes the form
\begin{equation}
    \widehat{\mathscr{G}}_{\text{full}}[\alpha]=\widehat{\mathscr{G}}[\alpha]-\hbar\kappa^{-1}\sum_{p\in\EuScript{P}_{\gamma}}\left(\widehat{\mathcal{E}}[\alpha]-\frac{i\kappa k }{2\pi}\widehat{F}[\alpha]\right)(p)
    \label{Gauss:fullTheory3}
\end{equation}
with $\widehat{\mathscr{G}}[\alpha]$ the Gauss constraint operator acting on the bulk Hilbert space given by \eqref{DH.eq:3.14}. Assuming that the smearing function $\alpha$ vanishes on the boundary, the full constraint operator \eqref{Gauss:fullTheory} reduces to the bulk Gauss constraint $\widehat{\mathscr{G}}[\alpha]$ implying gauge-invariance of the quantum state in the bulk. As a consequence, from \eqref{Gauss:fullTheory3}, one obtains the additional constraint equation    
\begin{equation}
\mathds{1}\otimes\widehat{F}_{\underline{A}}(p)=-\frac{2\pi i}{\kappa k}\widehat{\mathcal{E}}_{\underline{A}}(p)\otimes\mathds{1}   
\label{Gauss:fullTheory4}
\end{equation}
at each puncture $p\in\EuScript{P}_{\gamma}$. Note that, by definition, $\widehat{\mathcal{E}}_{\underline{A}}(p)$ can be related to the quantized super electric flux via $\widehat{\mathcal{E}}_{\underline{A}}(p)=\lim_{\epsilon\rightarrow 0}\widehat{\mathcal{X}}_{\underline{A}}(D_{\epsilon})$ and thus, according to \eqref{DH.eq:1.27}, acts in terms of right- resp. left-invariant vector fields. Hence, from \eqref{Gauss:fullTheory4}, we deduce that the Hilbert space of the quantized boundary degrees of freedom corresponds to the Hilbert space of a quantized super Chern-Simons theory on $\Delta$ with punctures $\EuScript{P}_{\gamma}$. This leads to the well-known (super)conformal blocks. In the pure bosonic theory, these play an important role in the context of the computation of the black hole entropy.

As already outlined above, in \cite{Ghosh:2014rra}, an alternative route in describing the entropy of black hole has been studied. More precisely, assuming that the edges piercing the boundary are of infinitesimal but nonzero width, this induces infinitesimal holes localized at the punctures on the boundary which then gives rise to new physical degrees of freedom that are localised at the corner $\partial\Delta$.

In the following, let us describe these new degrees of freedom in the context of chiral supergravity. To this end, generalizing the discussion in \cite{Geiller:2017xad} in context of the bosonic theory to the super category, let us consider the following quantities defined on the canonical phase space of the super Chern-Simons theory
\begin{equation}
    \mathcal{O}[\alpha]:=-\frac{k}{2\pi}\int_{\Delta}\braket{\alpha F(\mathcal{A}^+)}+\frac{k}{2\pi}\int_{\partial\Delta}\braket{\alpha\mathcal{A}^+}=\frac{k}{2\pi}\int_{\Delta}\braket{\mathrm{d}\alpha\wedge\mathcal{A}^+-\frac{1}{2}\alpha[\mathcal{A}^+\wedge\mathcal{A}^+]}
    \label{D.eq:1.4.17}
\end{equation}
where $\alpha$ denotes an arbitrary $\mathrm{Lie}(\mathcal{G})$-valued smearing function on $\Delta$. In case that $\alpha$ vanishes on the corner, this quantity reduces to the flatness constraint \eqref{D.eq:1.4.16}, i.e., $\mathcal{O}[\alpha]\equiv\mathcal{F}[\alpha]$ if $\alpha|_{\partial\Delta}=0$. Computing the graded Poisson bracket between $\mathcal{O}[\alpha]$ and the super connection, one finds 
\begin{equation}
    \{\mathcal{O}[\alpha],\mathcal{A}_a^{+\underline{A}}\}=D^{(\mathcal{A}^+)}_a\alpha^{\underline{A}}
    \label{D.eq:1.4.18}
\end{equation}
This is in fact immediate to see using \eqref{D.eq:1.4.15}. For instance, direct calculation yields
\begin{equation}
    \{\frac{k}{2\pi}\int_{\Delta}\,\braket{\mathrm{d}\alpha\wedge\mathcal{A}^+},\mathcal{A}_a^{+\underline{A}}(x)\}=\frac{k}{2\pi}\int_{\Delta}\mathrm{d}^2y\,\epsilon^{bc}\mathscr{S}_{\underline{C}\underline{B}}\partial_b\alpha^{\underline{B}}(y)\{\mathcal{A}_c^{+\underline{C}}(y),\mathcal{A}_a^{+\underline{A}}(x)\}=\partial_a\alpha^{\underline{A}}(x)
    \label{D.eq:1.4.19}
\end{equation}
On the other hand, one has
\begin{equation}
    -\frac{k}{4\pi}\{\int_{\Delta}\,\braket{\alpha[\mathcal{A}^+\wedge\mathcal{A}^+]},\mathcal{A}_a^{+\underline{A}}(x)\}=[\mathcal{A}^+_a,\alpha]^{\underline{A}}(x)
    \label{D.eq:1.4.20}
\end{equation}
which, together with \eqref{D.eq:1.4.19}, directly gives \eqref{D.eq:1.4.18}. With these preparations, let us next compute the Poisson algebra among the $\mathcal{O}[\alpha]$. Using identity \eqref{D.eq:1.4.18}, it follows for arbitrary smearing functions $\alpha$ and $\beta$ that
\begin{align}
    \{\mathcal{O}[\alpha],\mathcal{O}[\beta]\}&=\frac{k}{2\pi}\int_{\Delta}(-1)^{|\alpha||\beta|}\braket{\mathrm{d}\beta\wedge D^{(\mathcal{A}^+)}\alpha-\beta[\mathcal{A}^+\wedge D^{(\mathcal{A}^+)}\alpha])}\nonumber\\
    &=-\frac{k}{2\pi}\int_{\Delta}\braket{D^{(\mathcal{A}^+)}\alpha\wedge D^{(\mathcal{A}^+)}\beta}
    \label{D.eq:1.4.21}
\end{align}
Since $D^{(\mathcal{A}^+)}D^{(\mathcal{A}^+)}\beta=[F(\mathcal{A}^+),\beta]$, one has
\begin{align}
    \braket{D^{(\mathcal{A}^+)}\alpha\wedge D^{(\mathcal{A}^+)}\beta}&=\mathrm{d}\!\braket{\alpha D^{(\mathcal{A}^+)}\beta}-\braket{\alpha[F(\mathcal{A}^+),\beta]}\nonumber\\
    &=\braket{\mathrm{d}\alpha\wedge\mathrm{d}\beta}-\mathrm{d}\!\braket{[\alpha,\beta]\mathcal{A}^+}+\braket{[\alpha,\beta]F(\mathcal{A}^+)}
    \label{D.eq:1.4.22}
\end{align}
Thus, inserting \eqref{D.eq:1.4.22} into \eqref{D.eq:1.4.21} and assuming that $\alpha$ is vanishes on the corner $\partial\Delta$, it follows 
\begin{align}
    \{\mathcal{F}[\alpha],\mathcal{O}[\beta]\}&=\mathcal{F}[[\alpha,\beta]]\simeq 0
    \label{D.eq:1.4.23}
\end{align}
where we used that $[\alpha,\beta]|_{\partial\Delta}=0$. Thus, it follows that $\mathcal{O}[\alpha]$ weakly Poisson commutes with the flatness constraint. That is, $\mathcal{O}[\alpha]$ defines a weak \emph{Dirac observable}. Moreover, for smearing functions $\alpha$ and
and $\beta$ with $\alpha|_{\partial\Delta}=\beta|_{\partial\Delta}$, one has 
\begin{equation}
    \mathcal{O}[\alpha]-\mathcal{O}[\beta]=\mathcal{O}[\alpha-\beta]\equiv-\mathcal{F}[\alpha-\beta]\simeq 0
    \label{D.eq:1.4.24}
\end{equation}
Hence, it follows that the observables $\mathcal{O}[\alpha]$ are localized on the corner. Furthermore, by \eqref{D.eq:1.4.21} and \eqref{D.eq:1.4.22}, they satisfy the following graded Poisson relations
\begin{align}
    \{\mathcal{O}[\alpha],\mathcal{O}[\beta]\}&=\mathcal{O}[[\alpha,\beta]]+\frac{k}{2\pi}\int_{\partial\Delta}\braket{\mathrm{d}\alpha,\beta}
    \label{D.eq:1.4.25}
\end{align}
Since, the last term on the right-hand side of Equation \eqref{D.eq:1.4.25} is completely field-independent, it, in particular, Poisson commutes with all the corner observables $\mathcal{O}[\alpha]$. Thus, it follows that the Poisson algebra among the $\mathcal{O}[\alpha]$ is indeed closed up to a central term. 

In this context, recall that, given an Abelian (bosonic) Lie algebra $\mathfrak{a}$, a \emph{central extension} of a super Lie algebra $\mathfrak{g}$ (not necessarily finite-dimensional) by $\mathfrak{a}$ is defined as a short exact sequence \cite{Kenji:2000}
\begin{equation}
    0\rightarrow\mathfrak{a}\rightarrow\mathfrak{h}\stackrel{\pi}{\rightarrow}\mathfrak{g}\rightarrow 0
\end{equation}
with $\mathfrak{h}$ a super Lie algebra such that $[\mathfrak{a},\mathfrak{h}]=0$ and $\pi:\,\mathfrak{h}\rightarrow\mathfrak{g}$ an even surjective super Lie algebra morphism yielding the identification $\mathfrak{h}/\mathfrak{a}\cong\mathfrak{g}$.

In our concrete situation, at each puncture, $\partial\Delta$ is topologically equivalent to a 1-dimensional circle. Thus, in this case, it follows that a basis of smearing functions $\alpha$ is given by functions $\alpha_N^{\underline{A}}$ of the form
\begin{equation}
    \alpha_N^{\underline{A}}|_{\partial\Delta}:=e^{i N\theta}T^{\underline{A}},\quad\alpha_N^{\underline{A}}|_{\Delta\setminus\partial\Delta}\equiv 0
    \label{D.eq:1.4.26}
\end{equation}
where $\theta\in[0,2\pi]$ denotes the angle coordinate parametrizing the circle, $N\in\mathbb{Z}$ and $(T^{\underline{A}})_{\underline{A}}$ is a homogeneous basis of $\mathfrak{osp}(\mathcal{N}|2)_{\mathbb{C}}$. From \eqref{D.eq:1.4.25}, it then follows that the corresponding corner observables $q^{\underline{A}}_N:=\mathcal{O}[\alpha_N^{\underline{A}}]$ satisfy the Poisson relations
\begin{equation}
    \{q^{\underline{A}}_M,q^{\underline{B}}_N\}=\tensor{f}{^{\underline{A}\underline{B}}_{\underline{C}}}q^{\underline{C}}_{M+N}+N\delta_{M+N,0}(T^{\underline{A}},T^{\underline{B}})
    \label{D.eq:1.4.27}
\end{equation}
where $(T^{\underline{A}},T^{\underline{B}}):=ik\braket{T^{\underline{A}},T^{\underline{B}}}$ and $\tensor{f}{^{\underline{A}\underline{B}}_{\underline{C}}}$ denote the structure coefficients defined via 
\begin{equation}
[T^{\underline{A}},T^{\underline{B}}]=\tensor{f}{^{\underline{A}\underline{B}}_{\underline{C}}}T^{\underline{C}}    
\label{D.eq:1.4.28}
\end{equation}
Interestingly, \eqref{D.eq:1.4.28} are precisely the graded commutation relations of a \emph{Kac-Moody superalgebra} corresponding to the affinisation of $\mathfrak{osp}(\mathcal{N}|2)_{\mathbb{C}}$ \cite{Kenji:2000}. It follows via the so-called \emph{Sugawara construction}, that the generators of the Kac-Moody superalgebra can be used in order to generate representations of the \emph{super Virasoro algebra} \cite{Goddard:1986ee}. Thus, to conclude, the singularities induced by the intersection of super spin networks with the boundary give rise to new physical degrees of freedom living on the corner which are associated to \emph{superconformal field theories} and which, in analogy to \cite{Ghosh:2014rra} in context of the bosonic theory, may also account for black hole entropy and hence may play a role in the quantum description of supersymmetric black holes in the framework of LQG.

\section{Continuous representations of $\mathrm{OSp}(1|2)$ and the reality of the super area operator}
\label{sec:ospreps}
In this section, we would like to derive a certain class of infinite-dimensional representations of $\mathrm{OSp}(1|2)_{\mathbb{C}}$. This is motivated by the observation that, according to \eqref{eq:3.3.49}, the super area operator of chiral LQSG, in general, has complex eigenvalues. In fact, in case that the edges of a super spin network state are labeled by isospin quantum numbers $j\in\frac{\mathbb{N}_0}{2}$ corresponding to finite-dimensional irreducible representations of $\mathrm{OSp}(1|2)_{\mathbb{C}}$, the eigenvalue of the super area operator becomes purely imaginary.\\
This is in complete analogy to the bosonic theory. In \cite{Frodden:2012dq,Achour:2014eqa}, in the context of the self-dual theory, it has been observed that, in order to obtain physically realistic (real) eigenvalues for the standard area operator with $\beta=-i$, the edges of the spin network states necessarily have to be labeled by spin quantum numbers $j\in\mathbb{C}$ corresponding to certain infinite-dimensional irreducible representations of $\mathrm{SL}(2,\mathbb{C})$.\\
In what follows, in Section \ref{section 4.1}, we first would like to review the so-called principal series of the real form $\mathrm{SL}(2,\mathbb{R})$. As we will see, their corresponding complexifications indeed provide the (unique) subclass of irreducible representations of $\mathrm{SL}(2,\mathbb{C})$ as studied for instance in \cite{Frodden:2012dq,Achour:2014eqa} with respect which the standard area operator of LQG for $\beta=-i$ becomes purely real.\\   
Subsequently, in Section \ref{section 4.2}, we will study a generalization of these kind of representation to irreducible representations of the corresponding super Lie group $\mathrm{OSp}(1|2)$. We will then demonstrate that their corresponding complexifications contain a (unique) subclass of representations of $\mathrm{OSp}(1|2)_{\mathbb{C}}$ that lead to a physically realistic super area operator.\\
\\
Before we proceed, however, let us first introduce a suitable basis of the super Lie algebra $\mathfrak{osp}(1|2)_{\mathbb{C}}$. As summarized in the Appendix \ref{Appendix:Supergroups}, $\mathfrak{osp}(\mathcal{N}|2)_{\mathbb{C}}$ is generated by the homogeneous basis $(T_i^+,Q^r_{A},T^{rs})$ with $i\in\{1,2,3\}$, $A\in\{\pm\}$ and $r=1,\ldots,\mathcal{N}$ satisfying the graded commutation relations \eqref{eq:C11a}-\eqref{eq:C11d}. In the case $\mathcal{N}=1$, one can arrive at a Cartan-Weyl basis $(J_3,J_{\pm},V_{\pm})$ of the superalgebra by setting
\begin{align}
J_{\pm}:=-i(T_{1}^{+}\pm i T_{2}^{+}),\quad  J_3:=iT_{3}^{+},\quad V_{\pm}:=\pm\frac{\sqrt{L}}{2}(i-1)Q_{\pm}
\end{align}
It then follows from Eq. \eqref{eq:C11a} that the commutators among the even generators satisfy
\begin{equation}
[J_3,J_{\pm}]=\pm J_{\pm},\quad [J_+,J_-]=2J_3
\label{AppendixA.1:commutation1}
\end{equation}
which are the standard commutation relations of $\mathfrak{sl}(2,\mathbb{R})$. For the remaining commutators, it follows
\begin{alignat}{3}
&[J_3,V_{\pm}]=\pm\frac{1}{2}V_{\pm},\quad && [J_{\mp},V_{\pm}]=V_{\mp},\quad && [J_{\pm},V_{\pm}]=0\\
&[V_{\pm},V_{\pm}]=\pm\frac{1}{2}J_{\pm},\quad && [V_+,V_-]=-\frac{1}{2}J_3
\end{alignat}
These are the standard commutation relations of the corresponding real form $\mathrm{OSp}(1|2)$ that we will use in what follows.

\subsection{Review: Principal series representations of $\mathrm{SL}(2,\mathbb{R})$}\label{section 4.1}
In the following, let us review the so-called principal series representations of $\mathrm{SL}(2,\mathbb{R})$. To this end, we will follow the references \cite{Vogan:81,Vogan:2020}. \\
The principal series representations of $\mathrm{SL}(2,\mathbb{R})$ can be derived from the highly reducible representation $(\pi,W)$ of $\mathrm{SL}(2,\mathbb{R})$ on the space $W:=C^{\infty}(\mathbb{R}^2\setminus\{0\})$ of smooth functions on the punctured plane given by
\begin{equation}
    (\pi(g)f)(v):=f(g^{-1}v)
    \label{eq:4.1}
\end{equation}
$\forall v\in\mathbb{R}^2\setminus\{0\}$ by restricting onto the proper subsets 
\begin{equation}
    W_{j}^{\epsilon}:=\{f\in W|\,f(tx)=t^{2j}f(x)\,\forall t>0\wedge f(-x)=(-1)^{\epsilon}f(x)\}
    \label{eq:4.2}
\end{equation}
with $j\in\mathbb{C}$ an arbitrary complex number and parity $\epsilon\in\mathbb{Z}_2$. By the homogeneity property, it follows that $W_j^{\epsilon}$ can be identified with a certain subclass of smooth functions on the unit circle, i.e., 
\begin{equation}
    W_{j}^{\epsilon}\cong C^{\infty}_{j}(\mathbb{S}^1)^{\epsilon}
    \label{eq:4.3}
\end{equation}
In what follows, let us choose a global chart of $\mathbb{S}^1$ via
\begin{equation}
    [0,4\pi)\ni\theta\mapsto e^{i\frac{\theta}{2}}\in\mathbb{S}^1
    \label{eq:4.4}
\end{equation}
The representation $\pi$ induces a corresponding pushforward representation $\pi_*$ of $\mathfrak{sl}(2,\mathbb{R})$ on $W_j^{\epsilon}$ via
\begin{equation}
    (\pi_*(X)f)(v):=\frac{\mathrm{d}}{\mathrm{d}s}\bigg{|}_{s=0}f(e^{-sX}v)
    \label{eq:4.5}
\end{equation}
$\forall X\in\mathfrak{sl}(2,\mathbb{R})$ and $v\in\mathbb{R}^2\setminus\{0\}$. In this way, one obtains explicit expressions for the representations $\widehat{J}_3:=\pi_{*}(J_3)$ and $\widehat{J}_{\pm}:=\pi_{*}(J_{\pm})$ of the generators $(J_{\pm},J_3)$ w.r.t. the chosen global chart \eqref{eq:4.4}. However, the form of these operators turn out to be less suitable for further computations. Therefore, one constructs a new representation for the $(J_{\pm},J_3)$ by replacing $\widehat{J}_3\rightarrow\frac{1}{2i}(\widehat{J}_+-\widehat{J}_-)$ and $\widehat{J}_{\pm}\rightarrow\widehat{J}_3\pm\frac{i}{2}(\widehat{J}_++\widehat{J}_-)$. In this way, one finds that
\begin{align}
    \widehat{J}_3=-i\partial_{\theta},\quad \widehat{J}_{\pm}=e^{\pm i\theta}\partial_{\theta}\mp ie^{\pm i\theta}j
    \label{eq:4.6}
\end{align}
As can be verified by direct computation, the operators \eqref{eq:4.6} indeed satisfy the commutation relations \eqref{AppendixA.1:commutation1}. Via the identification \eqref{eq:4.3}, the vector space $W_j^{\epsilon}$ contains vectors of the form $w_m:=e^{im\theta}$ with $m\in\frac{\mathbb{Z}}{2}$ and $2m\equiv\epsilon\,\mathrm{mod}\,2$, i.e., $m$ is a proper integer or half-integer, respectively, depending on whether $\epsilon$ is even or odd. Let $V_j^{\epsilon}\subset W_j^{\epsilon}$ be defined as the algebraic span of the $w_m$. By equipping $W_j^{\epsilon}$ (resp. $V_j^{\epsilon}$) with a suitable topology, such as a Hilbert space topology induced by the unique invariant Haar measure on $\mathbb{S}^1$, it follows that $V_j^{\epsilon}$ is dense in $W_j^{\epsilon}$ (see also \cite{Vogan:81,Vogan:2020} for more details). Using the explicit expressions \eqref{eq:4.6}, it follows that 
\begin{align}
\widehat{J}_3 w_m=m w_m,\quad\widehat{J}_{\pm}w_m=i(m\mp j)w_{m\pm 1}
\label{eq:4.7}
\end{align}
Hence, from \eqref{eq:4.7} we deduce that for the case where $j\in\mathbb{C}$ is neither an integer nor half-integer, the restriction of the representation $\pi_*$ to $V_j^{\epsilon}$ is irreducible. On the other hand, if $j\in\frac{\mathbb{Z}}{2}$, the representation is reducible and by taking sums and intersection of suitable subsets one can construct (finite-dimensional) irreducible subspaces which we also denote by $V_j^{\epsilon}$ and which lead to the well-known finite-dimensional spin-$j$ representations of the corresponding compact real form $\mathrm{su}(2)$. Finally, let us compute the quadratic Casimir operator $C_2$ of $\mathfrak{sl}(2,\mathbb{R})$ given by 
\begin{align}
    \widehat{C}_2=&(\widehat{J}_3)^2+\frac{1}{2}(\widehat{J}_+\widehat{J}_-+\widehat{J}_-\widehat{J}_+)=(\widehat{J}_3)^2+\widehat{J}_3+\widehat{J}_-\widehat{J}_+\nonumber\\
    =&\left(\widehat{J}_3+\frac{1}{2}\right)^2+\widehat{J}_-\widehat{J}_+-\frac{1}{4}
    \label{eq:4.8}
\end{align}
on the irreducible subspaces $V_j^{\epsilon}$. Applying \eqref{eq:4.8} on the vectors $w_m$ and using \eqref{eq:4.7}, we find
\begin{align}
    \widehat{C}_2w_m&=\left[\left(m+\frac{1}{2}\right)^2-(m-j)(m+1+j)-\frac{1}{4}\right]w_m\nonumber\\
    &=j(j+1)w_m
    \label{eq:4.9}
\end{align}
so that 
\begin{equation}
    \widehat{C}_2=j(j+1)\mathds{1}
    \label{eq:4.10}
\end{equation}
Hence, as expected, the quadratic Casimir operator on $V_j^{\epsilon}$ is an integer multiply of the identity operator on $V_j^{\epsilon}$.
\begin{remark}
Defining the operator $\widehat{\Omega}:=4\widehat{C}_2+1$ it follows from \eqref{eq:4.10} that  
\begin{equation}
    \widehat{\Omega}=(2j+1)^2\mathds{1}
    \label{eq:4.11}
\end{equation}
Hence, the irreducible representations $(\pi|_{V_j^{\epsilon}},V_j^{\epsilon})$ correspond the principal series representations $(\pi_{\lambda},V_{\lambda})$ as defined in \cite{Vogan:81} labeled by $\lambda:=2j+1$.
\end{remark}

\begin{remark}
The complexification of the principal series representations of $\mathrm{SL}(2,\mathbb{R})$ derived above contains a sublcass of irreducible representations of $\mathrm{SL}(2,\mathbb{C})$ that lead to a physically consistent area operator in LQG using self-dual variables. In fact, for $\beta=-i$, the eigenvalues of the standard area operator of LQG are of the form 
\begin{equation}
    -8\pi i l_p^2\sqrt{j(j+1)}
    \label{eq:4.12}
\end{equation}
Hence, according to \eqref{eq:4.12}, it follows that the eigenvalues become real iff the quadratic Casimir operator is negative definite which is the case for instance if $\widehat{\Omega}$ is negative definite. By \eqref{eq:4.11}, this is only the case if $j=-\frac{1}{2}+is$ for some real number $s\in\mathbb{R}$. This leads back to the continuous series of $\mathrm{SL}(2,\mathbb{R})$ as studied in \cite{Frodden:2012dq,Achour:2014eqa} in the context of the black hole entropy computation in the self-dual theory.
\end{remark}

\subsection{Principal series representations of $\mathrm{OSp}(1|2)$}\label{section 4.2}
With these preliminaries, we would like to derive continuous representations of the super Lie group $\mathrm{OSp}(1|2)$ which are graded generalizations of the principal series representations of the underlying bosonic subgroup $\mathrm{SL}(2,\mathbb{R})$. The finite-dimensional irreducible representations of $\mathfrak{osp}(1|2)$ are well-known (see for instance \cite{Scheunert:1976wj,Scheunert:1976wi,Minnaert:1990sz,Berezin:1981}). In \cite{Hadasz:2013bwa}, a certain class of continuous representations of the quantum group $\mathrm{OSp}_q(1|2)$ with $q\in\mathbb{S}^1$ has been given. However, to the best of the authors' knowledge, continuous representations for the super Lie group $\mathrm{OSp}(1|2)$, so far, have not been studied in the literature.\\
To this end, in the following, we will derive an explicit series of representations (both finite- and inifinite-dimensional) of the corresponding super Lie algebra $\mathfrak{osp}(1|2)$ and then use the Super Harish-Chandra Theorem (see \cite{Tuynman:2004,Tuynman:2018}) in order to lift these representations to representations of the super Lie group $\mathrm{OSp}(1|2)$. To do so, let us first state the following useful observation.
\begin{prop}\label{Prop:4.3}
Let $\pi_*:\,\mathfrak{osp}(1|2)\rightarrow\mathrm{Op}(\mathcal{D},\mathfrak{H})$ be a linear map from the super Lie algebra $\mathfrak{osp}(1|2)$ to the space $\mathrm{Op}(\mathcal{D},\mathfrak{H})$ of (un)bounded operators on a super Hilbert space $\mathfrak{H}$ mutually defined on some dense graded subspace $\mathcal{D}\subset\mathfrak{H}$. Then, $\pi_*$ defines a representation of $\mathfrak{osp}(1|2)$, i.e., a morphism of super Lie algebras iff it satisfies the identities
\begin{equation}
    [\pi_*(J_3),\pi_*(V_{\pm})]=\pm\frac{1}{2}\pi_{*}(V_{\pm})\text{ and }[\pi_*(V_+),\pi_*(V_-)]=-\frac{1}{2}\pi_*(J_3)
    \label{eq:4.2.1}
\end{equation}
as well as 
\begin{equation}
    \pi_{*}(J_{\pm})=\pm 4\pi_{*}(V_{\pm})^2
    \label{eq:4.2.2}
\end{equation}
\end{prop}

\begin{remark}\label{Remark:4.4}
Prop. \ref{Prop:4.3} states that, given an even operator $\widehat{J}_{3}$ as well as odd operators $\widehat{V}_{\pm}$ on a super Hilbert space $\mathfrak{H}$ mutually defined on some dense graded subspace $\mathcal{D}\subset\mathfrak{H}$, these operators can be associated to a representation of the super Lie algebra $\mathfrak{osp}(1|2)$ provided that they satisfy the relations
\begin{equation}
[\widehat{J}_3,\widehat{V}_{\pm}]=\pm\frac{1}{2}\widehat{V}_{\pm},\quad[\widehat{V}_+,\widehat{V}_-]=-\frac{1}{2}\widehat{J}_3
\label{eq:4.2.3}
\end{equation}
The last identity \eqref{eq:4.2.2} can merely be interpreted as a defining equation for the representations of the remaining bosonic generators $J_{\pm}$ by setting $\widehat{J}_{\pm}:=\pm 4\widehat{V}_{\pm}^2$.
\end{remark}

\begin{proof}[Proof of Prop. 4.2]
One direction is immediate, so suppose that $\pi_*:\,\mathfrak{osp}(1|2)\rightarrow\mathrm{Op}(\mathcal{D},\mathfrak{H})$ is a linear map satisfying the relations \eqref{eq:4.2.1} as well as \eqref{eq:4.2.2}. Since $\mathrm{Op}(\mathcal{D},\mathfrak{H})$ defines a super Lie algebra, the graded Jacobi identity holds on $\mathrm{Op}(\mathcal{D},\mathfrak{H})$. Let $\widehat{J}_{\pm}:=\pi_{*}(J_{\pm})$ and similar for the other generators. Then, by the graded Jacobi identity, it follows
\begin{align}
    [\widehat{J}_{+},\widehat{J}_-]&=-4\left[[\widehat{V}_+,\widehat{V}_+],[\widehat{V}_-,\widehat{V}_-]\right]\nonumber\\
    &=4\left[\widehat{V}_-,\left[\widehat{V}_-,[\widehat{V}_+,\widehat{V}_+]\right]\right]-4\left[\widehat{V}_-,\left[[\widehat{V}_+,\widehat{V}_+],\widehat{V}_-\right]\right]\nonumber\\
    &=-8\left(\left[\widehat{V}_-,\left[\widehat{V}_+,[\widehat{V}_+,\widehat{V}_-]\right]\right]+\left[\widehat{V}_-,\left[\widehat{V}_+,[\widehat{V}_-,\widehat{V}_+]\right]\right]\right)\nonumber\\
    &=8\left[\widehat{V}_-,[\widehat{V}_+,\widehat{J}_3]\right]\nonumber\\
    &=-4[\widehat{V}_-,\widehat{V}_+]=2\widehat{J}_3
    \label{eq:4.2.4}
\end{align}
On the other, one finds
\begin{align}
    [\widehat{J}_3,\widehat{J}_{\pm}]&=\pm\left[\widehat{J}_3,[\widehat{V}_{\pm},\widehat{V}_{\pm}]\right]\nonumber\\
    &=\mp 2\left(\left[\widehat{V}_{\pm},[\widehat{V}_{\pm},\widehat{J}_{3}]\right]-\left[\widehat{V}_{\pm},[\widehat{J}_3,\widehat{V}_{\pm}]\right]\right)\nonumber\\
    &=\pm 4\left[\widehat{V}_{\pm},[\widehat{J}_{3},\widehat{V}_{\pm}]\right]\nonumber\\
    &=2[\widehat{V}_{\pm},\widehat{V}_{\pm}]=\pm\widehat{J}_{\pm}
    \label{eq:4.2.5}
\end{align}
Therefore, the bosonic operators indeed define a representation of the Lie algebra $\mathfrak{sl}(2,\mathbb{R})$. The remaining commutators can be shown similarly.
\end{proof}
Hence, according to Prop. \ref{Prop:4.3} and Remark \ref{Remark:4.4}, it suffices to guess explicit expressions for the operators $\widehat{V}_{\pm}$ and $\widehat{J}_3$ and subsequently check whether the identities \eqref{eq:4.2.1} are indeed satisfied.\\
By restriction, any irreducible representation of $\mathfrak{osp}(1|2)$ induces a (possibly) reducible representation of the corresponding bosonic sub Lie algebra $\mathfrak{sl}(2,\mathbb{R})$ which itself may be decomposable into the irreps as stated in the previous section. Therefore, as the underlying super vector space $\mathcal{V}$ of such a representation, let us propose
\begin{equation}
    \mathcal{V}=V_j^{\epsilon}\otimes\Pi V_{j'}^{\epsilon'}
    \label{eq:4.2.6}
\end{equation}
with $V_j^{\epsilon}$ (resp. $V_{j'}^{\epsilon'}$) as defined in section \ref{section 4.1}.\footnote{Since the theory of (un)operators on super Hilbert spaces seems to be not that well-explored (but see \cite{}), we will keep the following discussion purely algebraic and discuss algebraic representations of super Lie algebras in terms of (un)bounded operators on super Vector spaces without specifying the topology and identifying them as dense subspaces of larger super Hilbert spaces.} On this super vector space, we then define the operators
\begin{equation}
    \widehat{V}_+=\begin{pmatrix}
    0 & e^{i\frac{\theta}{2}}\\
    e^{i\frac{\theta}{2}}\partial_{\theta}-ie^{i\frac{\theta}{2}}j & 0
    \end{pmatrix},\quad \widehat{V}_-=\frac{i}{4}\begin{pmatrix}
    0 & e^{-i\frac{\theta}{2}}\\
    e^{-i\frac{\theta}{2}}\partial_{\theta}+ie^{-i\frac{\theta}{2}}j & 0
    \end{pmatrix}
    \label{eq:4.2.7}
\end{equation}
as well as
\begin{equation}
    \widehat{J}_3:=-i\begin{pmatrix}
    \partial_{\theta} & 0\\
    0 & \partial_{\theta}
    \end{pmatrix}
    \label{eq:4.2.8}
\end{equation}
From this, it is immediate to see that
\begin{equation}
    [\widehat{J}_3,\widehat{V}_{\pm}]=\pm\frac{1}{2}\widehat{V}_{\pm}
    \label{eq:4.2.9}
\end{equation}
On the other hand, by direct computation, one finds
\begin{equation}
    [\widehat{V}_+,\widehat{V}_-]=\frac{i}{4}\begin{pmatrix}
    2\partial_{\theta} & 0\\
    0 & 2\partial_{\theta}
    \end{pmatrix}=-\frac{1}{2}\widehat{J}_3
    \label{eq:4.2.10}
\end{equation}
Thus, the identities \eqref{eq:4.2.1} are satisfied and the operators \eqref{eq:4.2.7} as well as \eqref{eq:4.2.8} can indeed be associated to representations of $\mathfrak{osp}(1|2)$. The remaining bosonic generators of $\mathfrak{sl}(2,\mathbb{R})$ are given by 
\begin{equation}
    \widehat{J}_+=4\widehat{V}_+^2=4\begin{pmatrix}
    e^{i\theta}\partial_{\theta}-ie^{i\theta}j & 0\\
    0 & e^{i\theta}\partial_{\theta}-ie^{i\theta}\left(j-\frac{1}{2}\right)
    \end{pmatrix}
    \label{eq:4.2.11}
\end{equation}
and 
\begin{equation}
    \widehat{J}_-=-4\widehat{V}_-^2=\frac{1}{4}\begin{pmatrix}
    e^{-i\theta}\partial_{\theta}+ie^{-i\theta}j & 0\\
    0 & e^{-i\theta}\partial_{\theta}+ie^{-i\theta}\left(j-\frac{1}{2}\right)
    \end{pmatrix}
    \label{eq:4.2.12}
\end{equation}
respectively. By comparing with the discussion in section \ref{section 4.1}, we thus infer that the odd part $V_{j'}^{\epsilon'}$ of the super vector space $\mathcal{V}$ can be identified with $V_{j'}^{\epsilon'}=V_{j-\frac{1}{2}}^{\epsilon+1}$ so that
\begin{equation}
    \mathcal{V}_j^{\epsilon}\equiv\mathcal{V}=V_{j}^{\epsilon}\oplus\Pi V_{j-\frac{1}{2}}^{\epsilon+1}
    \label{eq:4.2.13}
\end{equation}
By construction, it is clear that these representations which we would like to denote by $(\pi_j,\mathcal{V}_j^{\epsilon})$ in what follows are irreducible. For $j\in\frac{\mathbb{Z}}{2}$, these are isomorphic to the finite-dimensional representations of $\mathrm{OSp}(1|2)$ as discussed in \cite{Scheunert:1976wj,Scheunert:1976wi,Minnaert:1990sz,Berezin:1981}. That these representations are indeed irreducible can also be checked by computing the quadratic Casimir operator $\widehat{C}_2^{\mathfrak{osp}}$ given by 
\begin{equation}
    \widehat{C}_2^{\mathfrak{osp}}=\widehat{C}_2+\widehat{V}_+\widehat{V}_--\widehat{V}_-\widehat{V}_+=\widehat{C}_2+\frac{1}{2}\widehat{J}_3+2\widehat{V}_+\widehat{V}_-
    \label{eq:4.2.14}
\end{equation}
with $\widehat{C}_2$ the quadratic Casimir operator of the bosonic subalgebra $\mathfrak{sl}(2,\mathbb{R})$ defined via \eqref{eq:4.8}. By definition, this operator commutes with all generators of $\mathfrak{osp}(1|2)$. Therefore, since the representations are irreducible, by a supersymmetric generalization of Schur's Lemma, it has to be a scalar multiple of the identity operator $\mathds{1}$ on $\mathcal{V}_j^{\epsilon}$.\\
To see see that this is indeed the case, note that a basis of the super vector space $\mathcal{V}$ is provided by states of the form $(w_m,0)^T$ and $(0,w_n)^T$ with $m,n\in\frac{\mathbb{Z}}{2}$ satisfying $2m\equiv\,\epsilon\,\mathrm{mod}\,2$ and $2n\equiv\,\epsilon+1\,\mathrm{mod}\,2$ (and possible additional restrictions on $m$ and $n$ in case $j$ is an (half-)integer to account for irreducibility). Using the explicit representations \eqref{eq:4.2.7}, we then find
\begin{align}
    \widehat{V}_+\widehat{V}_-\begin{pmatrix}w_m\\
    0
    \end{pmatrix}=-\frac{1}{4}(m+j)\widehat{V}_+\begin{pmatrix}0\\
    w_{m-\frac{1}{2}}
    \end{pmatrix}=-\frac{1}{4}(m+j)\begin{pmatrix}w_m\\
    0
    \end{pmatrix}
    \label{eq:4.2.15}
\end{align}
On the other hand, we have
\begin{align}
    \widehat{V}_+\widehat{V}_-\begin{pmatrix}0\\
    w_m
    \end{pmatrix}=\frac{i}{4}\widehat{V}_+\begin{pmatrix}w_{m-\frac{1}{2}}\\
    0
    \end{pmatrix}=-\frac{1}{4}\left(m-j-\frac{1}{2}\right)\begin{pmatrix}0\\
    w_m
    \end{pmatrix}
    \label{eq:4.2.16}
\end{align}
Thus, from \eqref{eq:4.2.15} and \eqref{eq:4.2.16} we obtain
\begin{align}
    \widehat{C}_2^{\mathfrak{osp}}\begin{pmatrix}w_m\\
    0
    \end{pmatrix}=\left[j(j+1)+\frac{m}{2}-\frac{1}{2}(m+j)\right]\begin{pmatrix}w_m\\
    0
    \end{pmatrix}=j\left(j+\frac{1}{2}\right)\begin{pmatrix}w_m\\
    0
    \end{pmatrix}
    \label{eq:4.2.17}
\end{align}
as well as
\begin{align}
    \widehat{C}_2^{\mathfrak{osp}}\begin{pmatrix}0\\
    w_m
    \end{pmatrix}=\left[\left(j-\frac{1}{2}\right)\left(j+\frac{1}{2}\right)+\frac{m}{2}-\frac{1}{2}\left(m-j-\frac{1}{2}\right)\right]\begin{pmatrix}0\\
    w_m
    \end{pmatrix}=j\left(j+\frac{1}{2}\right)\begin{pmatrix}0\\
    w_m
    \end{pmatrix}
    \label{eq:4.2.18}
\end{align}
Thus, to summarize, we find
\begin{align}
    \widehat{C}_2^{\mathfrak{osp}}=j\left(j+\frac{1}{2}\right)\mathds{1}
    \label{eq:4.2.19}
\end{align}
that is, the quadratic Casimir operator is a scalar multiply of the identity operator with proportionality factor given by $j(j+\frac{1}{2})$.\\
Having derived a graded generalization of the principal series representations, let us finally show that this series indeed contains a subclass of irreducible representations with respect to which the super area operator as studied in section \ref{Section:chSUGRA-superspin} becomes purely real and thus physically consistent. Therefore, recall that, according to \eqref{graded area with Casimir}, the eigenvalues of the super area operator become real iff the quadratic Casimir operator $\widehat{C}_2^{\mathfrak{osp}}$ is negative definite. This is the case for instance if the corresponding operator  

\begin{align}
    \widehat{\Omega}^{\mathfrak{osp}}:=\widehat{C}_2^{\mathfrak{osp}}+\frac{1}{16}=\left(j+\frac{1}{4}\right)^2\mathds{1}
    \label{eq:4.2.20}
\end{align}
becomes negative definite which in turn is the case iff
\begin{align}
    \widehat{\Omega}^{\mathfrak{osp}}\leq 0\Leftrightarrow j\in-\frac{1}{4}+i\mathbb{R}
    \label{eq:4.2.21}
\end{align}
For $j=-\frac{1}{4}+is$ with $s\in\mathbb{R}$, this then yields
\begin{align}
    \widehat{C}_2^{\mathfrak{osp}}=\left(is-\frac{1}{4}\right)\left(is+\frac{1}{4}\right)\mathds{1}=-\left(s^2+\frac{1}{16}\right)\mathds{1}
    \label{eq:4.2.22}
\end{align}
Hence, according to \eqref{eq:3.3.49}, in this case it follows that the action of the super area operator takes the form
\begin{equation}
\widehat{\mathrm{gAr}}(S) T_{\gamma,\Vec{\pi},\Vec{m},\Vec{n}}=8\pi l_p^2\sqrt{s^2+\frac{1}{16}}\,T_{\gamma,\Vec{\pi},\Vec{m},\Vec{n}}
\label{eq:4.2.23}
\end{equation}
That is, super spin network states whose edges are labeled by isospin quantum numbers $j$ satisfying \eqref{eq:4.2.21} are indeed eigenstates of the super area operator with real eigenvalues. Interestingly, this is in complete analogy to the bosonic theory.

\section{Entropy calculation}\label{sec:EntropyBH}
\subsection{Super characters of $\mathrm{UOSp(1|2)}$ and the Verlinde formula}
As discussed in Section \ref{OutlookBHentropy}, the boundary theory of chiral loop quantum supergravity for the case $\mathcal{N}=1$ is described by a super Chern-Simons theory with punctures and gauge supergroup $\mathrm{OSp}(1|2)_{\mathbb{C}}$ as well as complex Chern-Simons level. Hence, to the boundary one can associate an entropy in terms of the number of Chern-Simons degrees of freedom generated by the super spin network edges piercing the boundary. Unfortunately, the (super) Chern-Simons theory with complex and non-compact gauge group is not well-known. Moreover, it is not clear how to deal with the fact that the Chern-Simons level is purely imaginary. Interestingly, similar issues also seem to arise in the context of boundary theories in string theory \cite{Mikhaylov:2014aoa}.

In the following, we therefore want to adapt the strategy of \cite{Achour:2014eqa} in the context of the purely bosonic theory to the supersymmetric setting by studying a specific compact real form of $\mathrm{OSp}(1|2)_{\mathbb{C}}$ and then performing an analytic continuation to the corresponding complex Lie supergroup. More precisely, in what follows, let us consider the Chern-Simons theory with compact gauge supergroup $\mathrm{UOSp}(1|2)$ and integer Chern-Simons level $k=-12\pi/\kappa\Lambda_{\mathrm{cos}}$ and punctures labeled by finite-dimensional irreducible representations $\{\Vec{j}\}$ of $\mathrm{UOSp}(1|2)$ with $j\in\frac{\mathbb{N}_0}{2}$. Let us then compute the number $\mathcal{N}_k(\{\Vec{j}\})$ of Chern-Simons degrees of freedom given by the dimension of the superconformal blocks. We then finally perform an analytic continuation by replacing $j\rightarrow j=-\frac{1}{4}+is$ for some $s\in\mathbb{R}$ for each $j\in\{\Vec{j}\}$ as well as $k\rightarrow ik$ in $\mathcal{N}_k(\{\Vec{j}\})$. Moreover, in order to simplify the discussion, we assume that the boundary $H$ is topologically of the form $\mathbb{R}\times\mathbb{S}^2$, that is, the $2$-dimensional slices $\Delta_t$ are topologically equivalent to $2$-spheres. Furthermore, let us consider the limit $k\rightarrow\infty$ corresponding to a vanishing cosmological constant $\Lambda_{\mathrm{cos}}$. Under these assumptions, it follows that the number of microstates $\mathcal{N}_{\infty}(\{\Vec{j}\})$  is given by the number of $\mathrm{UOSp}(1|2)$ gauge-invariant states, i.e., it can be identified with the number of trivial subrepresentations contained in the tensor product representation $\bigotimes_j \pi_j$.

In what follows, in view of the analytic continuation, we want to derive an explicit integral formula for $\mathcal{N}_{\infty}(\{\Vec{j}\})$. To this end, following \cite{Berezin:1981}, let us first review some important facts unitary orthosymplectic group $\mathrm{UOSp}(1|2)$.

On the complex Grassmann algebra $\Lambda^{\mathbb{C}}$, we introduce a conjugation rule that is parity preserving and satisfies
\begin{align}
    \bar{\alpha\beta}&=\bar{\alpha}\bar{\beta},\quad\bar{c\alpha}=\bar{c}\bar{\alpha}\label{eq:5.1.1}\\
    \bar{\bar{\alpha}}&=(-1)^{|\alpha|}\alpha
    \label{eq:5.1.2}
\end{align}
for any homogeneous $\alpha,\beta\in\Lambda^{\mathbb{C}}$ and where $\bar{c}$ for a complex number $c\in\mathbb{C}$ denotes the usual complex conjugation. This can be extended to an adjointness relation on the whole super Lie module $\mathfrak{osp}(1|2)_{\mathbb{C}}$  by setting
\begin{equation}
    (T_i^+)^{\dagger}:=-T_i^+,\quad (V_{\pm})^{\dagger}=\pm V_{\mp}
    \label{eq:5.1.3}
\end{equation}
In this way, the (real) unitary orthosymplectic Lie superalgebra $\mathfrak{uosp}(1|2)$ can be defined as the superalgebra 
\begin{equation}
    \mathfrak{uosp}(1|2):=\{X\in\mathfrak{osp}(1|2)_{\mathbb{C}}|\,X^{\dagger}=-X\}
    \label{eq:5.1.4}
\end{equation}
A general even element $X\in\mathfrak{uosp}(1|2)_0$ can be expanded in the form $X=\alpha^iT_i^++\bar{\eta}V_++\eta V_-$ with Grassmann-even $\alpha^i\in\Lambda_0^{\mathbb{C}}$ and odd $\eta\in\Lambda_1^{\mathbb{C}}$. By the Super Harish-Chandra Theorem (see \cite{Tuynman:2004}), the super Lie group $\mathrm{UOSp}(1|2)$ can be identified with the globally split supermanifold\footnote{Here, we identify $\mathrm{SU}(2)$ with the corresponding purely bosonic split supermanifold $\mathbf{S}(\mathrm{SU}(2))$ (see \cite{Eder:2022})} $\mathrm{SU}(2)\times\mathfrak{uosp}(1|2)_0$  via the canonical isomorphism
\begin{align}
    \Phi:\,\mathrm{SU}(2)\times\mathfrak{uosp}(1|2)_0&\rightarrow\mathrm{UOSp}(1|2)\label{eq:5.1.5}\\
    (g,\bar{\eta}V_++\eta V_-)&\mapsto g\cdot\exp(\bar{\eta}V_++\eta V_-)\nonumber
\end{align}
where, using Euler coordinates $(\phi,\vartheta,\psi)$ for $\mathrm{SU}(2)\cong\mathbb{S}^3$, a general group element  $g\in\mathrm{SU}(2)\subset\mathrm{UOSp}(1|2)$ of the underlying bosonic subgroup can be expanded in the form 
\begin{equation}
    g=\exp(\varphi T_3^+)\exp(\vartheta T_2^+)\exp(\psi T_3^+)
    \label{eq:5.1.6}
\end{equation}
Using the identification \eqref{eq:5.1.5}, it is easy to compute the invariant integral on $\mathrm{UOSp}(1|2)$ (see \cite{Berezin:1981} as well as \cite{Eder:2022,Tuynman:2018}). After some short calculation, one finds \cite{Berezin:1981}
\begin{equation}
    \int_{\mathrm{UOSp}(1|2)}=\int_{\mathbb{S}^3}\mathrm{d}\sigma\int_B\mathrm{d}\bar{\eta}\,\mathrm{d}\eta\,\left(1-\frac{1}{4}\bar{\eta}\eta\right)
    \label{eq:5.1.7}
\end{equation}
with $\mathrm{d}\sigma$ given by 4 times the normalized volume form on the three-sphere $\mathbb{S}^3$ which in Euler coordinates reads 
\begin{equation}
    \mathrm{d}\sigma=\frac{1}{4\pi^2}\sin\vartheta\,\mathrm{d}\varphi\,\mathrm{d}\vartheta\,\mathrm{d}\psi
    \label{eq:5.1.8}
\end{equation}
The reason for choosing this normalization is that one then has
\begin{equation}
    \int_{\mathrm{UOSp}(1|2)}\,\mathds{1}=1
    \label{eq:5.1.9}
\end{equation}
for the constant unit function $\mathds{1}$ on $\mathrm{UOSp}(1|2)$. using the invariant integral \eqref{eq:5.1.7}, we can define a super scalar product $\mathscr{S}$ on $H^{\infty}(\mathrm{UOSp}(1|2),\mathbb{C})\cong C^{\infty}(\mathrm{SU}(2))\otimes\Lambda^{\mathbb{C}}_2$ via
\begin{equation}
    \mathscr{S}(f|h):=\int_{\mathrm{UOSp}(1|2)}\bar{f}h
    \label{eq:5.1.10}
\end{equation}
for any $f,h\in H^{\infty}(\mathrm{UOSp}(1|2),\mathbb{C})$ where $\bar{f}$ is computed using the conjugation rule as defined above (see Eqs. \eqref{eq:5.1.1}-\eqref{eq:5.1.2}).  

Next, we want to discuss the characters on the unitary orthosymplectic supergroup. As shown in \cite{Berezin:1981}, the finite-dimensional irreducible representations $\pi_j$ of $\mathrm{UOSp}(1|2)$ with $j\in\frac{\mathbb{N}_0}{2}$ acquire the following matrix representation 
\begin{equation}
    \pi_j=\left(
    \begin{array}{c;{2pt/2pt}c}
        T^{j}_{j\,m_1,j\,m_2} & T^{j}_{j\,m_1,j-\frac{1}{2}\,m_2} \\ \hdashline[2pt/2pt]
    T^{j}_{j-\frac{1}{2}\,m_1,j\,m_2} & T^{j}_{j-\frac{1}{2}\,m_1,j-\frac{1}{2}\,m_2}
    \end{array}
\right)
\label{eq:5.1.11}
\end{equation}
with matrix coefficients given by the explicit formulas 
\begin{equation}
    T^{j}_{j\,m_1,j\,m_2}(g,\eta,\bar{\eta})=\left(1-\frac{j}{4}\bar{\eta}\eta\right)D^j_{m_1,m_2}(g)
    \label{eq:5.1.12}
\end{equation}
and
\begin{equation}
    T^{j}_{j-\frac{1}{2}\,m_1,j-\frac{1}{2}\,m_2}(g,\eta,\bar{\eta})=\left(1+\frac{j+\frac{1}{2}}{4}\bar{\eta}\eta\right)D^{j-\frac{1}{2}}_{m_1,m_2}(g)
    \label{eq:5.1.13}
\end{equation}
as well as 
\begin{equation}
    T^{j}_{j-\frac{1}{2}\,m_1,j\,m_2}(g,\eta,\bar{\eta})=-\frac{1}{2}\sqrt{j-m_2}\bar{\eta}D^{j-\frac{1}{2}}_{m_1,m_2+\frac{1}{2}}(g)+\frac{1}{2}\sqrt{j+m_2}\eta D^{j-\frac{1}{2}}_{m_1,m_2-\frac{1}{2}}(g)
    \label{eq:5.1.14}
\end{equation}
and 
\begin{equation}
    T^{j}_{j\,m_1,j-\frac{1}{2}\,m_2}(g,\eta,\bar{\eta})=-\frac{1}{2}\sqrt{j+m_2+\frac{1}{2}}\bar{\eta}D^{j}_{m_1,m_2+\frac{1}{2}}(g)-\frac{1}{2}\sqrt{j-m_2+\frac{1}{2}}\eta D^{j}_{m_1,m_2-\frac{1}{2}}(g)
    \label{eq:5.1.15}
\end{equation}
where $D^j_{m_1,m_2}$ denotes the matrix coefficients of the spin-$j$ representations of $\mathrm{SU}(2)$. Using the super scalar product \eqref{eq:5.1.10}, one then computes
\begin{align}
    \mathscr{S}(T^{j}_{j\,m_1,j\,m_2}|T^{j'}_{j'\,m'_1,j'\,m'_2})&=\braket{\braket{D^j_{m_1,m_2}|D^{j'}_{m'_1,m'_2}}}\int_B\mathrm{d}\bar{\eta}\,\mathrm{d}\eta\,\left(1-\frac{1}{4}\bar{\eta}\eta\right)\left(1-\frac{j+j'}{4}\bar{\eta}\eta\right)\nonumber\\
    &=\frac{4}{2j+1}\delta_{j j'}\delta_{m_1 m'_1}\delta_{m_2 m'_2}\frac{1}{4}(2j+1)\nonumber\\
    &=\delta_{j j'}\delta_{m_1 m'_1}\delta_{m_2 m'_2}
    \label{eq:5.1.16}
\end{align}
where $\braket{\braket{\cdot,\cdot}}$ denotes the positive-definite scalar product on $\mathrm{SU}(2)$ induced by $\mathrm{d}\sigma$. On the other hand, one finds
\begin{align}
    \mathscr{S}(T^{j}_{j-\frac{1}{2}\,m_1,j-\frac{1}{2}\,m_2}|T^{j'}_{j'-\frac{1}{2}\,m'_1,j'-\frac{1}{2}\,m'_2})&=\frac{4}{2j}\delta_{j j'}\delta_{m_1 m'_1}\delta_{m_2 m'_2}\int_B\mathrm{d}\bar{\eta}\,\mathrm{d}\eta\,\left(1-\frac{1}{4}\bar{\eta}\eta\right)\left(1+\frac{2j+1}{4}\bar{\eta}\eta\right)\nonumber\\
    &=-\delta_{j j'}\delta_{m_1 m'_1}\delta_{m_2 m'_2}
    \label{eq:5.1.17}
\end{align}
Furthermore, for the super scalar product between mixed matrix coefficients, it follows
\begin{align}
    \mathscr{S}(T^{j}_{j\,m_1,j\,m_2}|T^{j'}_{j'-\frac{1}{2}\,m'_1,j'-\frac{1}{2}\,m'_2})&=\frac{4}{2j+1}\delta_{j,j'-\frac{1}{2}}\delta_{m_1 m'_1}\delta_{m_2 m'_2}\int_B\mathrm{d}\bar{\eta}\,\mathrm{d}\eta\,\left(1-\frac{1}{4}\bar{\eta}\eta\right)\left(1+\frac{1}{4}\bar{\eta}\eta\right)\nonumber\\
    &=0
    \label{eq:5.1.18}
\end{align}
that is, they are orthogonal. Taking a closer look at Eqs. \eqref{eq:5.1.16} and \eqref{eq:5.1.17} one observes that these are normalized up to a relative minus sign. In fact, this seems to be in contradiction to the results of \cite{Berezin:1981}. However, as we will see explicitly below, this relative minus sign will be crucial to prove the orthogonality of the super characters on $\mathrm{UOSp}(1|2)$. Finally, for the remaining matrix coefficients, one computes
\begin{align}
    \mathscr{S}(T^{j}_{j-\frac{1}{2}\,m_1,j\,m_2}|T^{j'}_{j'-\frac{1}{2}\,m'_1,j'\,m'_2})=-\delta_{j j'}\delta_{m_1 m'_1}\delta_{m_2 m'_2}
    \label{eq:5.1.19}
\end{align}
as well as 
\begin{align}
    \mathscr{S}(T^{j}_{j\,m_1,j-\frac{1}{2}\,m_2}|T^{j'}_{j'\,m'_1,j'-\frac{1}{2}\,m'_2})=-\delta_{j j'}\delta_{m_1 m'_1}\delta_{m_2 m'_2}
    \label{eq:5.1.20}
\end{align}
with all the remaining combinations being zero. With these preliminaries, we are ready to introduce the super characters on the supergroup $\mathrm{UOSp}(1|2)$. The super character $\chi_j$ associated to the finite-dimensional irreducible representation $\pi_j$ of $\mathrm{UOSp}(1|2)$ is a smooth map $\chi_j:\,\mathrm{UOSp}(1|2)\rightarrow\Lambda^{\mathbb{C}}_0$ defined via
\begin{equation}
    \chi_j(g):=\mathrm{str}(\pi_j(g)),\quad\forall g\in\mathrm{UOSp}(1|2)
    \label{eq:5.1.21}
\end{equation}
Using the matrix representation \eqref{eq:5.1.11}, it can explicitly be written in the form
\begin{equation}
    \chi_j=\sum_{m=-j}^j{T^j_{j\,m,j\,m}}-\sum_{m=-j+\frac{1}{2}}^{j-\frac{1}{2}}{T^j_{j-\frac{1}{2}\,m,j-\frac{1}{2}\,m}}
    \label{eq:5.1.22}
\end{equation}
By the Eqs. \eqref{eq:5.1.16}-\eqref{eq:5.1.18}, it then follows immediately from \eqref{eq:5.1.22} that the super scalar product between two super characters $\chi_j$ and $\chi_{j'}$ is given by 
\begin{align}
    \mathscr{S}(\chi_j|\chi_{j'})&=\sum_{m,m'}\mathscr{S}(T^j_{j\,m,j\,m}|T^{j'}_{j'\,m',j'\,m'})+\sum_{m,m'}\mathscr{S}(T^j_{j-\frac{1}{2}\,m,j-\frac{1}{2}\,m}|T^{j'}_{j'-\frac{1}{2}\,m',j'-\frac{1}{2}\,m'})\nonumber\\
    &=(2j+1)\delta_{j j'}-2j\delta_{j j'}\nonumber\\
    &=\delta_{j j'}
    \label{eq:5.1.23}
\end{align}
that is, the super characters are normalized and two distinct super characters associated to two inequivalent irreducible representation are pairwise orthogonal. It is clear that one can associate a super character to any (not necessarily irreducible) finite-dimensional representation of $\mathrm{UOSp}(1|2)$. By definition, it then follows that super characters are well-behaved under (graded) tensor product and direct sum such that, for instance, $\chi_{j\otimes j'}=\chi_j\cdot\chi_{j'}$ and $\chi_{j\oplus j'}=\chi_j+\chi_{j'}$. 

By exploiting these properties, we are now able to derive an explicit integral formula for $\mathcal{N}_{\infty}(\{\Vec{j}\})$. To do so, for later purposes, let us subdivide $\{\Vec{j}\}$ into $p\leq n$ subfamilies $(n_l,j_l)$, $l=1,\ldots,p$, consisting of $0<n_l\leq n$ punctures labeled by $j_l\in\{\Vec{j}\}$. By the orthonomality property of the super characters, it follows that the number of Chern-Simons degrees of freedom on the boundary in the limit $k\rightarrow\infty$ is given by 
\begin{equation}
    \mathcal{N}_{\infty}(\{n_l,j_l\}_l)=\mathscr{S}\left(\chi_0\bigg{|}\prod_{l=1}^p(\chi_{j_l})^{n_l}\right)=\int_{\mathrm{UOSp}}\,\prod_{l=1}^p(\chi_{j_l}(\theta))^{n_l}
    \label{eq:5.1.24}
\end{equation}
We want to re-express \eqref{eq:5.1.24} in an even more explicit way. To this end, note that, according to \eqref{eq:5.1.22} as well as \eqref{eq:5.1.12} and \eqref{eq:5.1.13}, the super character $\chi_j$ can be written in the form
\begin{equation}
    \chi_j=\left(1-\frac{j}{4}\bar{\eta}\eta\right)\chi_j^{\mathrm{SU}(2)}-\left(1+\frac{j+\frac{1}{2}}{4}\bar{\eta}\eta\right)\chi_{j-\frac{1}{2}}^{\mathrm{SU}(2)}
    \label{eq:5.1.25}
\end{equation}
with $\chi_j^{\mathrm{SU}(2)}$ the character associated to the spin-$j$ representation of $\mathrm{SU}(2)$. This can also rewritten as follows
\begin{align}
    \chi_j=&\chi_j^{\mathrm{SU}(2)}-\chi_{j-\frac{1}{2}}^{\mathrm{SU}(2)}-\frac{1}{4}\bar{\eta}\eta\left(j\chi_j^{\mathrm{SU}(2)}+\left(j+\frac{1}{2}\right)\chi_{j-\frac{1}{2}}^{\mathrm{SU}(2)}\right)\nonumber\\
    =&\chi_j^{\mathrm{SU}(2)}-\chi_{j-\frac{1}{2}}^{\mathrm{SU}(2)}-\frac{1}{4}\bar{\eta}\eta\left(\left(\left(j+\frac{1}{4}\right)-\frac{1}{4}\right)\chi_j^{\mathrm{SU}(2)}+\left(\left(j+\frac{1}{4}\right)+\frac{1}{4}\right)\chi_{j-\frac{1}{2}}^{\mathrm{SU}(2)}\right)\nonumber\\
    =&\left(1+\frac{1}{16}\bar{\eta}\eta\right)\left(\chi_j^{\mathrm{SU}(2)}-\chi_{j-\frac{1}{2}}^{\mathrm{SU}(2)}\right)-\frac{1}{4}\left(j+\frac{1}{4}\right)\bar{\eta}\eta\left(\chi_j^{\mathrm{SU}(2)}+\chi_{j-\frac{1}{2}}^{\mathrm{SU}(2)}\right)\nonumber\\
    =&\left(1+\frac{1}{16}\bar{\eta}\eta\right)\left(\chi_j^{\mathrm{SU}(2)}-\chi_{j-\frac{1}{2}}^{\mathrm{SU}(2)}\right)-\frac{d_j}{16}\bar{\eta}\eta\left(\chi_j^{\mathrm{SU}(2)}+\chi_{j-\frac{1}{2}}^{\mathrm{SU}(2)}\right)
    \label{eq:5.1.26}
\end{align}
with $d_j=4j+1$ the (ungraded) dimension of the representation $\pi_j$. Using the explicit formula $\chi_j^{\mathrm{SU}(2)}(\theta)=\frac{\sin((2j+1)\theta)}{\sin\theta}$, one obtains the following useful identities
\begin{equation}
    \chi^{\mathrm{SU}(2)}_j(\theta)-\chi^{\mathrm{SU}(2)}_{j-\frac{1}{2}}(\theta)=\frac{\cos(2(j+\frac{1}{4})\theta)}{\cos(\frac{\theta}{2})}=\frac{\cos(d_j\frac{\theta}{2})}{\cos(\frac{\theta}{2})}
    \label{eq:5.1.27}
\end{equation}
and
\begin{equation}
    \chi^{\mathrm{SU}(2)}_j(\theta)+\chi^{\mathrm{SU}(2)}_{j-\frac{1}{2}}(\theta)=\frac{\sin(2(j+\frac{1}{4})\theta)}{\sin(\frac{\theta}{2})}=\frac{\sin(d_j\frac{\theta}{2})}{\sin(\frac{\theta}{2})}
    \label{eq:5.1.28}
\end{equation}
By reinserting into \eqref{eq:5.1.26}, this yields
\begin{align}
    \chi_j=\left(1+\frac{1}{16}\bar{\eta}\eta\right)\frac{\cos(d_j\frac{\theta}{2})}{\cos(\frac{\theta}{2})}-\frac{d_j}{16}\frac{\sin(d_j\frac{\theta}{2})}{\sin(\frac{\theta}{2})}\bar{\eta}\eta
    \label{eq:5.1.29}
\end{align}
This is a quite useful formula which will play a central for the computation of the entropy in what follows. For $N\in\mathbb{N}_0$, it gives
\begin{align}
    (\chi_j)^N=\left(1+\frac{N}{16}\bar{\eta}\eta\right)\left(\frac{\cos(d_j\frac{\theta}{2})}{\cos(\frac{\theta}{2})}\right)^N-N\frac{d_j}{16}\frac{\sin(d_j\frac{\theta}{2})}{\sin(\frac{\theta}{2})}\left(\frac{\cos(d_j\frac{\theta}{2})}{\cos(\frac{\theta}{2})}\right)^{N-1}\bar{\eta}\eta
    \label{eq:5.1.30}
\end{align}
Hence, the product of characters as appearing in the integral formula \eqref{eq:5.1.24} can be expanded in the form
\begin{align}
    \prod_{l=1}^p(\chi_{j_l})^{n_l}=&\left(1+\frac{n}{16}\bar{\eta}\eta\right)\prod_{l=1}^p\left(\frac{\cos(d_{j_l}\frac{\theta}{2})}{\cos(\frac{\theta}{2})}\right)^{n_l}\nonumber\\
    &-\frac{1}{16}\bar{\eta}\eta\sum_{l=1}^p n_l d_{j_l}\frac{\sin(d_{j_l}\frac{\theta}{2})}{\sin(\frac{\theta}{2})}\left(\frac{\cos(d_{j_l}\frac{\theta}{2})}{\cos(\frac{\theta}{2})}\right)^{n_l-1}\prod_{i\neq l}\left(\frac{\cos(d_{j_i}\frac{\theta}{2})}{\cos(\frac{\theta}{2})}\right)^{n_i}
    \label{eq:5.1.31}
\end{align}
with $n=\sum_l n_l$ the total number of punctures on $\Delta$. By inserting \eqref{eq:5.1.31} into \eqref{eq:5.1.24} and performing the Berezin integral as well as the variable substitution $\theta\rightarrow 2\theta$, we thus finally arrive at the following complicated but beautiful integral formula for the total number of $\mathrm{UOSp}(1|2)$ Chern-Simons degrees of freedom on the boundary\footnote{Since the super characters on $\mathrm{UOSp}(1|2)$ define class functions, it follows that via the identification $\mathrm{SU}(2)\cong\mathbb{S}^3$ explicitly given by
\begin{equation}
    \mathrm{SU}(2)\ni\begin{pmatrix}
    a & \bar{b}\\
    -\bar{b} & \bar{a}
    \end{pmatrix}\mapsto (x^1,x^2,x^3,x^4)^T:=(\Re a,\Im a, \Re b, \Im b)^T\in\mathbb{S}^3
    \label{eq:5.1.32}
\end{equation}
the super characters, when restricted to the bosonic subgroup, only depend on the $x^1$-coordinate. By performing the variable substitution $x^1:=\cos\theta$, it then follows that the invariant integral on $\mathrm{SU}(2)$ of a function $f\equiv f(\cos\theta)$ takes the form \cite{Straumann:2013}
\begin{equation}
    \int_{\mathbb{S}^3}\mathrm{d}\sigma\,f=-\frac{2}{\pi^2}\int\mathrm{d}(\cos\theta)\,f(\cos\theta)\int_{\mathbb{R}^4}\mathrm{d}^4x\,\delta(\|x\|^2-1)\delta(x^1-\cos\theta)=\frac{4}{\pi}\int_{0}^{2\pi}\mathrm{d}\theta\,\sin^2(\theta)f(\cos\theta)
    \label{eq:5.1.33}
\end{equation}
}
\begin{align}
    \mathcal{N}_{\infty}(\{n_l,j_l\})=&\frac{2}{\pi}\left(1-\frac{n}{4}\right)\int_0^{\pi}\mathrm{d}\theta\,\sin^2(2\theta)\prod_{l=1}^p\left(\frac{\cos(d_{j_l}\theta)}{\cos\theta}\right)^{n_l}\nonumber\\
    &+\frac{1}{2\pi}\sum_{l=1}^p n_l d_{j_l}\int_0^{\pi}\mathrm{d}\theta\,\sin^2(2\theta)\frac{\sin(d_{j_l}\theta)}{\sin\theta}\left(\frac{\cos(d_{j_l}\theta)}{\cos\theta}\right)^{n_l-1}\prod_{i\neq l}\left(\frac{\cos(d_{j_i}\theta)}{\cos\theta}\right)^{n_i}
    \label{eq:VerlindeFormula}
\end{align}
This formula is the starting point for the calculation of the entropy to be discussed in the following sections.

\subsection{The monochromatic case}
Having derived an integral formula for the number of $\mathrm{UOSp}(1|2)$ Chern-Simons degrees of freedom in the limit $k\rightarrow\infty$, we next want to use this formula in order to compute the entropy associated to the boundary in the framework of chiral loop quantum supergravity. To this end, following and adapting the ideas of \cite{Achour:2014eqa} in the context of the purely bosonic theory, we perform an analytic continuation of the Verlinde formula for the compact real form $\mathrm{UOSp}(1|2)$ to the corresponding non-compact complex gauge supergroup $\mathrm{OSp}(1|2)_{\mathbb{C}}$ of chiral LQSG by replacing the superspin quantum numbers $j\in\{\Vec{j}\}$ in \eqref{eq:VerlindeFormula} by $j\rightarrow -\frac{1}{4}+is$, i.e., quantum numbers corresponding to the principal series with respect to which the super area operator has purely real eigenvalues. 

To simplify the discussion, as a first step, in what follows let us focus on the monochromatic case and assume that the punctures on the boundary are all labeled by the same super spin quantum number $j$. Then, by replacing $j\rightarrow-\frac{1}{4}+is$ for some $s\in\mathbb{R}_{>0}$ in \eqref{eq:VerlindeFormula} for the special case $p=1$ and using $d_j=i4s=:i\tilde{s}$ as well as $\cos(ix)=\cosh(x)$ and $\sin(ix)=i\sinh(x)$, one finds that an analytically continued version of the Verlinde formula is given by the following integral formula
\begin{align}
  \mathcal{I}_{\infty}=&\frac{1}{\pi}\left(1-\frac{n}{4}\right)\oint_{\mathcal{C}}\mathrm{d}z\,\mu(z)\exp\left(n\ln\left(\frac{\cos(\tilde{s} z)}{\cosh z}\right)\right)\nonumber\\
    &-\frac{1}{4\pi} n\tilde{s}\oint_{\mathcal{C}}\mathrm{d}z\,\mu(z)\frac{\tan(\tilde{s}z)}{\tan z}\exp\left(n\ln\left(\frac{\cos(\tilde{s} z)}{\cosh z}\right)\right)
    \label{eq:5.2.1}
\end{align}
with density $\mu(z):=i\sinh^2(2z)$. Here, $\mathcal{C}$ denotes a closed contour in the complex plane going through all the (non-degenerate) critical points of the ``action''
\begin{equation}
    \mathcal{S}(z)=\ln\left(\frac{\cos(\tilde{s} z)}{\cosh z}\right)
    \label{eq:5.2.2}
\end{equation}
located along the imaginary axis and lying between 0 and $i\pi$. In what follows, we would like to evaluate the integral formula \eqref{eq:5.2.1} in the macroscopic limit corresponding to the limit $s\rightarrow\infty$ and $n\rightarrow\infty$, that is, large color as well as a large number of punctures on the boundary. In this limit, it then follows that we can apply the method of steepest descent. To this end, we need to determine the critical points of the action \eqref{eq:5.2.2}. Taking the complex derivative of \eqref{eq:5.2.2}, this gives 
\begin{equation}
    \frac{\mathrm{d}\mathcal{S}}{\mathrm{d}z}=-\tilde{s}\tan(\tilde{s}z)-\tanh(z)
    \label{eq:5.2.3}
\end{equation}
Thus, it follows that $z_c$ is a critical point of the action, i.e., $\mathcal{S}'(z_c)=0$ if and only if
\begin{equation}
    \tilde{s}\tan(\tilde{s}z_c)=-\tanh(z_c)
    \label{eq:5.2.4}
\end{equation}
If we restrict to critical points lying on the imaginary axis, it follows that, in the macroscopic limit, an approximate solution to Eq. \eqref{eq:5.2.4} is given by $z_c=i(\frac{\pi}{2}-\epsilon)$ for some small $\epsilon$ of order $\epsilon=o(\tilde{s}^{-1})$. Inserting this into the action \eqref{eq:5.2.2} and setting $\epsilon=\frac{1}{\tilde{s}}$, this gives 
\begin{align}
    \mathcal{S}(z_c)=&\ln\left(\frac{\cosh\left[\left(\frac{\pi}{2}-\epsilon\right)\tilde{s}\right]}{\cos\left(\frac{\pi}{2}-\epsilon\right)}\right)=\ln\left(\frac{e^{\frac{\pi}{2}\tilde{s}}}{2e\sin\epsilon}\right)\nonumber\\
    =&2\pi s+\ln\left(\frac{2s}{e}\right)
    \label{eq:5.2.5}
\end{align}
By using the identity \eqref{eq:5.2.4}, we find for the second derivative
\begin{equation}
    \mathcal{S}''(z_c)=-\tilde{s}^2(1+\tan^2(\tilde{s}z_c))-1+\tanh^2(z_c)\simeq -\tilde{s}^2
    \label{eq:5.2.6}
\end{equation}
Evaluation of the density $\mu(z)$ on the critical point gives
\begin{equation}
    \mu(z_c)\simeq\frac{1}{i\tilde{s}^2}
    \label{eq:5.2.7}
\end{equation}
Moreover, one finds
\begin{equation}
    \frac{\tan(\tilde{s}z_c)}{\tan z_c}\simeq -\frac{1}{\tilde{s}}
    \label{eq:5.2.8}
\end{equation}
Thus, in the macroscopic limit, it follows that the integral formula \eqref{eq:5.2.1} can be approximated by
\begin{align}
  \mathcal{I}_{\infty}=\frac{1}{\pi}\mu(z_c)\exp(n \mathcal{S}(z_c))\int_{-\infty}^{\infty}\mathrm{d}x\,\exp\left(-n \mathcal{S}''(z_c)\frac{x^2}{2}\right)
  \label{eq:5.2.9}
\end{align}
It is interesting to note that, due to \eqref{eq:5.2.8}, the term in the second line of \eqref{eq:5.2.1} cancels exactly with the second term in the first line proportional to the total number $n$ of punctures on the boundary. Thus, after performing the Gaussian integral, one finally ends up with
\begin{equation}
    \mathcal{I}_{\infty}=\sqrt{\frac{2}{\pi}}\frac{1}{64s^3\sqrt{n}}\left(\frac{2s}{e}\right)^n\exp\left(\frac{a_H}{4}-i\frac{\pi}{2}\right)
    \label{eq:5.2.10}
\end{equation}
with $a_H=8\pi ns$ the super area of the boundary in the monochromatic case (see Eq. \eqref{eq:4.2.23}). Interestingly, as we see, the analytic continuation of the state sum acquires an additional complex phase which seems to be counter intuitive. In fact, a similar observation has been made in the bosonic theory \cite{Achour:2014eqa}: There, the complex phase turned out to be even dependent on the number $n$ of punctures on the boundary. As a result, it has been suggested that one either has to consider the modulus of the analytically continued state sum formula or one needs to restrict to particular values for $n$ for which this additional complex phase vanishes. Since, here in the supersymmetric setting, the complex phase turns out to be in fact independent of the number of punctures, the first possibility seems to be most appropriate. This confirms the hypothesis of \cite{Achour:2014eqa} in the framework of the bosonic theory.

Taking the modulus of \eqref{eq:5.2.10}, we can immediately deduce that the leading order term for the entropy $S=\ln |\mathcal{I}_{\infty}|$ defined as the natural logarithm of the number of states is indeed given by Bekenstein-Hawking area law 
\begin{equation}
    S=\frac{a_H}{4l_p^2}+\ldots
    \label{eq:5.2.11}
\end{equation}
where we have have reintroduced physical units just for sake of clarity. This is a very intriguing result and follows here directly from the analytically continued Verlinde formula \eqref{eq:5.2.1}. In particular, we did not have to make any choices or fix the Barbero-Immirzi parameter to specific values. Moreover, this confirms the results of \cite{Achour:2014eqa} in the bosonic theory and supports the hypothesis that in the context of complex variables the entropy can be derived via an analytic continuation starting from a compact real form of the complex gauge group.

Let us finally determine the lower order quantum corrections to the entropy. To this end, as explained in detail in \cite{Achour:2014eqa}, note that the total number of punctures $n$ and the color $s$ both grow proportionally to $\sqrt{a_H}/l_p$. Hence, in the macroscopic limit, we can set $n=\nu\frac{\sqrt{a_H}}{l_p}$ as well as $s=\sigma\frac{\sqrt{a_H}}{l_p}$ with some numerical coefficients $\nu,\sigma>0$. Inserting this into \eqref{eq:5.2.10}, we then find that the entropy is given by 
\begin{equation}
    S=\ln |\mathcal{I}_{\infty}|=\frac{a_H}{4l_p^2}+\frac{\nu}{2}\frac{\sqrt{a_H}}{l_p}\ln\left(\frac{a_H}{l_p^2}\right)+\nu\ln\left(\frac{2\sigma}{e}\right)\frac{\sqrt{a_H}}{l_p}-\frac{7}{4}\ln\left(\frac{a_H}{l_p^2}\right)+\mathcal{O}(1)
    \label{eq:5.2.12}
\end{equation}
However, note that, so far, we have not taken into account the indistinguishability of punctures on the boundary. To do so, we have to divide out the total number of possible permutations of the punctures on the boundary, that is, we have to replace $|\mathcal{I}_{\infty}|$ by  $|\mathcal{I}_{\infty}|/n!$ in the formula of the entropy. In the macroscopic limit, we can approximate the number pf permutations $n!$ via the Stirling formula $n!\sim \sqrt{2\pi n}\left(\frac{n}{e}\right)^n$ which yields
\begin{equation}
    \ln n!=\frac{\nu}{2}\frac{\sqrt{a_H}}{l_p}\ln\left(\frac{a_H}{l_p^2}\right)+\nu\ln\left(\frac{\nu}{e}\right)\frac{\sqrt{a_H}}{l_p}+\frac{1}{4}\ln\left(\frac{a_H}{l_p^2}\right)+\mathcal{O}(1)
    \label{eq:5.2.13}
\end{equation}
Thus, if substract \eqref{eq:5.2.13} from \eqref{eq:5.2.12}, we then find that the effective entropy in the case of indistinguishable punctures is given by 
\begin{equation}
    S=\frac{a_H}{4l_p^2}+\nu\ln\left(\frac{2\sigma}{\nu}\right)\frac{\sqrt{a_H}}{l_p}-2\ln\left(\frac{a_H}{l_p^2}\right)+\mathcal{O}(1)
    \label{eq:5.2.14}
\end{equation}
Interestingly, this is almost the same formula as encountered in \cite{Achour:2014eqa} in the context of the bosonic theory. In particular, the logarithmic correction exactly coincides with the result of \cite{Achour:2014eqa}.

\subsection{The multi-color case}
So far, we have considered a simplified model assuming that the punctures on the boundary are all labeled by the same super spin quantum number. In this section, let us finally discuss the general case. To this end, following the same steps as in the previous section, we analytically continue the Verlinde formula \eqref{eq:VerlindeFormula} by replacing the super spin quantum numbers $j_l$ by $j_l\rightarrow-\frac{1}{4}+is_l$ with $s_l\in\mathbb{R}$ for $l=1,\ldots,p$, i.e., quantum numbers corresponding to principal series representations with respect to which the super area operator has purely real eigenvalues. In doing so, it follows, using $d_l=i4s_l=:i\tilde{s}_l$, that an analytically continued version of \eqref{eq:VerlindeFormula} is given by the following integral formula
\begin{align}
  \mathcal{I}_{\infty}=&\frac{1}{\pi}\left(1-\frac{n}{4}\right)\oint_{\mathcal{C}}\mathrm{d}z\,\mu(z)\exp\left(\sum_{l=1}^p n_l\ln\left(\frac{\cos(\tilde{s}_l z)}{\cosh z}\right)\right)\nonumber\\
    &-\frac{1}{4\pi}\sum_{l=1}^p n_l\tilde{s}_l\oint_{\mathcal{C}}\mathrm{d}z\,\mu(z)\frac{\tan(\tilde{s}_l z)}{\tan z}\exp\left(\sum_{i=1}^p n_i\ln\left(\frac{\cos(\tilde{s}_i z)}{\cosh z}\right)\right)
    \label{eq:5.3.1}
\end{align}
where, similar to the monochromatic case, $\mathcal{C}$ denotes a contour in the complex plane going through all the critical points of the ``action''
\begin{equation}
    \mathcal{S}(z):=\sum_{l=1}^p \nu_l\ln\left(\frac{\cos(\tilde{s}_l z)}{\cosh z}\right)
    \label{eq:5.3.2}
\end{equation}
Here and in what follows, we consider the macroscopic limit and assume that the number $n_l$ of punctures labeled by $j_l$ grow at same velocity so that $n_l=\kappa\nu_l$ for some large real number $\kappa\rightarrow\infty$ and some finite $\nu_l>0$ for $l=1,\ldots,p$. In this limit, we can again evaluate the integral formula \eqref{eq:5.3.1} by using the method of steepest decent. To this end, taking the complex derivative of \eqref{eq:5.3.2}, we find
\begin{equation}
    \frac{\mathrm{d}\mathcal{S}}{\mathrm{d}z}=-\sum_{l=1}^p \nu_l(\tilde{s}_l\tan(\tilde{s}_l z)+\tanh(z))
    \label{eq:5.3.3}
\end{equation}
Hence, it follows that critical points $z_c$ of the action \eqref{eq:5.3.2} are determined by the equation
\begin{equation}
    \sum_{l=1}^p \nu_l\tilde{s}_l\tan(\tilde{s}_l z)=-\left(\sum_{l=1}^p\nu_l\right)\tanh(z)
    \label{eq:5.3.4}
\end{equation}
If we again restrict to critical points lying along the imaginary axis, we find that in the macroscopic limit, i.e. $\kappa\rightarrow\infty$ and $s_l\rightarrow\infty$ $\forall l=1,\ldots,p$, an approximate solution to \eqref{eq:5.3.4} is given by $z_c=i(\frac{\pi}{2}-\epsilon)$ for some small $\epsilon$ of order $\epsilon=o(\bar{s}^{-1})$ with
\begin{equation}
    \bar{s}:=\frac{\sum_{l=1}^p\nu_l s_l}{\sum_{l=1}^p\nu_l}
    \label{eq:5.3.5}
\end{equation}
the \emph{mean color}. Inserting this into \eqref{eq:5.3.2} and setting $\epsilon=\frac{1}{4\bar{s}}$, this gives
\begin{align}
    \mathcal{S}(z_c):=&\sum_{l=1}^p \nu_l\ln\left(\frac{\cosh\left[\left(\frac{\pi}{2}-\epsilon\right)\tilde{s}_l\right]}{\cos\left(\frac{\pi}{2}-\epsilon\right)}\right)=\sum_{l=1}^p\nu_l\ln\left(\frac{e^{2\pi s_l}}{2e^{\frac{s_l}{\bar{s}}}\sin\epsilon}\right)\nonumber\\
    =&2\pi\sum_{l=1}^p \nu_l s_l+\sum_{l=1}^p\nu_l\ln\left(\frac{2\bar{s}}{e^{\frac{s_l}{\bar{s}}}}\right)=2\pi\sum_{l=1}^p \nu_l s_l+\ln\left((2\bar{s})^{\sum_{l=1}^p\nu_l} e^{-\frac{1}{\bar{s}}\sum_{l=1}^p\nu_l s_l}\right)\nonumber\\
    =&\frac{1}{\kappa}\frac{a_H}{4}+\frac{n}{\kappa}\ln\left(\frac{2\bar{s}}{e}\right)
    \label{eq:5.3.6}
\end{align}
with $a_H=8\pi\sum_{l=1}^p n_l s_l$ the super area of the boundary as measured with respect to the super area operator (see Eq. \eqref{eq:4.2.23}). Since $\sum_{l=1}^p n_l\tilde{s}_l=4n\bar{s}$ and $\frac{\tan(\tilde{s}z_c)}{\tan z_c}\simeq -\frac{1}{4\bar{s}}$, it then follows that, in the macroscopic limit, the analytically continued state sum formula \eqref{eq:5.3.1} takes the form
\begin{align}
  \mathcal{I}_{\infty}=&\frac{1}{\pi}\mu(z_c)\exp(\kappa \mathcal{S}(z_c))\int_{-\infty}^{\infty}\mathrm{d}x\,\exp\left(-\kappa \mathcal{S}''(z_c)\frac{x^2}{2}\right)
  \label{eq:5.3.7}
\end{align}
Again, it is interesting to note that, similar to the monochromatic case discussed in the previous section, the term in the second line of \eqref{eq:5.3.1} cancels exactly with the second term in the first line proportional to the total number $n$ of punctures on the boundary drastically simplifying the expression of the integral formula. Hence, by taking the modulus of \eqref{eq:5.3.7}, it follows immediately from \eqref{eq:5.3.6} that, at highest order, the entropy associated to the boundary is given by
\begin{equation}
    S=\ln|\mathcal{I}_{\infty}|=\frac{a_H}{4l_p^2}+\ldots
    \label{eq:5.3.8}
\end{equation}
and thus indeed corresponds to the Bekenstein-Hawking area law. The lower order quantum corrections can be computed similarly to the monochromatic case by replacing $s$ by the mean color $\bar{s}$.

\section{Discussion and outlook}
\label{sec:discussion}    
In this article we have shown that a large class of surfaces characterized by boundary conditions preserving local supersymmetry carry a surface theory with an entropy $S=A/4$  -- a quarter of its super-area in Planck units. This means that (a suitable generalization of) the Bekenstein-Hawking law holds in $\calN=1, D=4$ supergravity quantized with loop quantum gravity methods. 

There are several surprises that come together to yield this result:
The first is that the boundary theory and boundary conditions are uniquely fixed from the requirement of supersymmetry.
The second is that the boundary theory is a Chern-Simons theory and that it couples to the bulk just as for isolated horizons in the non-supersymmetric theory. 
The third is that there is a compact real form of $\Osp{1}{2}_{\mathbb{C}}$ that one can find a Verlinde-type formula for, that $\Osp{1}{2}_{\mathbb{C}}$ possesses representations with the right properties to carry out the analytic continuation prescription, and that the Verlinde formula allows it. Note also that due to the fact that the CS level is proportional to the inverse of the cosmological constant, the large-$k$ limit makes physical sense and one does not have to deal with the intricacies of \emph{quantum deformations} of super groups.   
Finally, the only change in comparison to the non-supersymmetric theory in highest order turns out to be a factor of 2 in the exponent which can be easily incorporated into the picture by using the area eigenvalue of two-sided punctures at the horizon.    

While the calculation proceeds in analogy with the one of \cite{Frodden:2012dq,Achour:2014eqa} for the bosonic case, there are also interesting differences: The class of surfaces admitted seems to be larger in our work. The Chern-Simons level of the boundary theory is not determined by geometric properties of the boundary, but by the cosmological constant. The quantity bounding the entropy is a supergeometric generalization of the area. Since there is  unbroken local supersymmetry on the boundary which takes the form of a gauge symmetry, bosonic area is simply not an observable in our context. It is not gauge invariant.  

There are several places where our arguments are not as stringent as they should be, and there are some open questions.
First of all, the bulk quantum theory is not complete, and the quantum theory for the boundary Chern-Simons theory for the non-compact supergroup and at imaginary level is not known directly. This is not satisfactory, but it is very similar to the situation for the non-supersymmetric theory in terms of chiral variables \cite{Frodden:2012dq,Achour:2014eqa}.  Moreover, we have not based our consideration on a theory of isolated horizons, since it has not been worked out yet for supergravity theories, as far as we know. Finally, the right-handed supersymmetry constraint has not been implemented in a direct way. Rather, our assumption is that it does not significantly reduce the number of surface states, as is assumed for the Hamilton constraint in the bosonic theory.  

The open issues mentioned above could all be starting points for future work. In addition, it would be interesting to extend the theory to extended supersymmetry, $\calN>1$. As we have pointed out in the previous sections, we have at least a good understanding of how the bulk theory would look like for $\calN=2$. Based on this it seems feasible to extend the entropy calculation to more physically realistic models with $\calN=2$. Complementary to this, it would be very desirable to complete a calculation for BPS black holes as considered in string theory \cite{Strominger:1996sh,Behrndt:1996jn}.

Finally, it is very interesting to note that $\Osp{m}{n}_\mathbb{C}$ super Chern-Simons theories at complex level $k$ show up as boundary theories in string theory \cite{Mikhaylov:2014aoa}. These theories are investigated in \cite{Mikhaylov:2014aoa} by intricate analytic continuation arguments starting from \emph{cs-supergroups}, which entail a choice of compact real form of the bosonic subgroup of the complex supergroup. It would be great to better understand the possible connections to the present work in general, and in particular to the analytic continuation we used.   

\section*{Acknowledgements}
We would like to thank Lee Smolin for communications at an early stage of this work and in particular for his interest in application of loop quantum gravity methods to supersymmetric black holes which was part of the motivation of this work. 
K.E. thanks the German Academic Scholarship Foundation (Studienstiftung des Deutschen Volkes) for financial support. H.S. acknowledges the contribution of the COST Action CA18108.


\appendix

\section{Super Chern-Simons theory}\label{Appendix:Chern-Simons}	
In this section, we want to briefly recall the basic definition and structure of the super Chern-Simons action. For more details on Chern-Simons theory with supergroup as a gauge group, we refer to \cite{Mikhaylov:2014aoa} as well as \cite{Cremonini:2019aao} studying the super Chern-Simons action in the geometric approach using integral forms.\\
Before we state the super Chern-Simons action, we need to introduce invariant inner products. Let $\mathcal{G}$ be a Lie supergroup. By the super Harish-Chandra theorem, the super Lie group has the equivalent characterization in terms of a \emph{super Harish-Chandra-pair} $(G,\mathfrak{g})$ with $G$ the underlying ordinary bosonic Lie group and $\mathfrak{g}$ the super Lie algebra of $\mathfrak{g}$ with $\mathfrak{g}_{\underline{0}}=\mathrm{Lie}(G)$\footnote{For the interested reader, we note that, for sake of concreteness, we will identify the (algebro-geometric) super Lie group with the corresponding Rogers-DeWitt supergroup using the functor of points prescription (see \cite{Eder:2022,Eder:2020erq} for more details).}.\\
A \emph{super metric} on $\mathfrak{g}$ is a bilinear map $\braket{\cdot,\cdot}:\,\mathfrak{g}\times\mathfrak{g}\rightarrow\mathbb{C}$ that is non-degenerate and graded-symmetric, i.e. $\braket{X,Y}=(-1)^{|X||Y|}\braket{Y,X}$ for any homogeneous $X,Y\in\mathfrak{g}$. Moreover, it is called Ad-\emph{invariant}, if 
\begin{equation}
\braket{\mathrm{Ad}_gX,\mathrm{Ad}_gY}=\braket{X,Y}\quad\forall g\in G
\label{eq:A1}
\end{equation}
and 
\begin{equation}
\braket{[Z,X],Y}+(-1)^{|X||Z|}\braket{X,[Z,Y]}=0
\label{eq:A2}
\end{equation}
for all homogeneous $X,Y,Z\in\mathfrak{g}$. This can be extended to a bilinear form $\braket{\cdot\wedge\cdot}:\,\Omega^p(\mathcal{M},\mathfrak{g})\times\Omega^q(\mathcal{M},\mathfrak{g})\rightarrow\Omega^{p+q}(\mathcal{M})$ on differential forms on a supermanifold $\mathcal{M}$ with values in the super Lie algebra $\mathfrak{g}$. Therefore, first note that the sheaf $\Omega^{\bullet}(\mathcal{M},\mathfrak{g})$ carries the structure of a $\mathbb{Z}\times\mathbb{Z}_2$-bigraded module, where, for any $\omega\in(\Omega^k(\mathcal{M}))_{\underline{i}}$, its parity $\epsilon(\omega)$ is defined as 
\begin{equation}
\epsilon(\omega):=(k,\underline{i})\in\mathbb{Z}\times\mathbb{Z}_2
\label{eq:}
\end{equation}
where we will also write $|\omega|:=\underline{i}$ for the underlying $\mathbb{Z}_2$-grading. For homogeneous $\mathfrak{g}$-valued differential forms $\omega\in\Omega^p(\mathcal{M},\mathfrak{g})$ and $\eta\in\Omega^q(\mathcal{M},\mathfrak{g})$, we then set
\begin{equation}
\braket{\omega\wedge\eta}:=(-1)^{|i|(|\eta|+|j|)}\omega^i\wedge\eta^j\braket{X_i,X_j}
\label{eq:A3}
\end{equation}
where we have chosen a real homogeneous basis $(X_i)_i$ of $\mathfrak{g}$ and simply wrote $|i|:=|X_i|$ for the parity. A direct calculation yields
\begin{align}
\braket{\omega\wedge\eta}&:=(-1)^{|i|(|\eta|+|j|)}\omega^i\wedge\eta^j\braket{X_i,X_j}\nonumber\\
&=(-1)^{pq}(-1)^{|i||\eta|}(-1)^{(|\omega|+|i|)(|\eta|+|j|)}\eta^j\wedge\omega^i\braket{X_j,X_i}\nonumber\\
&=(-1)^{pq}\braket{\eta\wedge\omega}
\label{eq:A4}
\end{align}
Finally, let us derive an important identity which plays a central role in may calculations. in fact, using the $\mathrm{Ad}$-invariance (\ref{eq:A2}), one obtains
\begin{align}
\braket{\omega\wedge[\eta\wedge\xi]}&=(-1)^{|i|(|\eta|+|\xi|+|j|+|k|)}(-1)^{|j|(|\xi|+|k|)}\omega^i\wedge\eta^j\wedge\xi^k\braket{X_i,[X_j,X_k]}\nonumber\\
&=(-1)^{|i|(|\eta|+|\xi|+|j|+|k|)}(-1)^{|j|(|\xi|+|k|)}\omega^i\wedge\eta^j\wedge\xi^k\braket{[X_i,X_j],X_k}\nonumber\\
&=(-1)^{|i|(|\eta|+|j|)}\braket{\omega^i\wedge\eta^j\otimes[X_i,X_j]\wedge\xi}\nonumber\\
&=\braket{[\omega\wedge\eta]\wedge\xi}
\label{eq:A5}
\end{align}
As discussed in Section \ref{sec:Review} (see also \cite{Eder:2021rgt} for more details), the Chern-Simons action naturally appears as a boundary term in chiral limit of the Holst-MacDowell-Mansouri action of supergravity. In fact, let $\mathcal{A}$ be a super connection and $F(\mathcal{A})$ its corresponding curvature, then
\begin{align}
\braket{F(\mathcal{A})\wedge F(\mathcal{A})}=\mathrm{d}\!\braket{\mathcal{A}\wedge F(\mathcal{A})-\frac{1}{6}\mathcal{A}\wedge[\mathcal{A}\wedge\mathcal{A}]}
\label{eq:A6}
\end{align}
To see this, note that
\begin{align}
\mathrm{d}\!\braket{\mathcal{A}\wedge F(\mathcal{A})-\frac{1}{6}\mathcal{A}\wedge[\mathcal{A}\wedge\mathcal{A}]}&=\braket{\mathrm{d}\mathcal{A}\wedge\mathrm{d}\mathcal{A}+\frac{1}{2}\mathrm{d}\mathcal{A}\wedge[\mathcal{A}\wedge\mathcal{A}]-\mathcal{A}\wedge[\mathrm{d}\mathcal{A}\wedge\mathcal{A}]}-\frac{1}{6}\mathrm{d}\!\braket{\mathcal{A}\wedge[\mathcal{A}\wedge \mathcal{A}]}\nonumber\\
&=\braket{\mathrm{d}\mathcal{A}\wedge\mathrm{d}\mathcal{A}+\frac{1}{3}\mathrm{d}\mathcal{A}\wedge[\mathcal{A}\wedge \mathcal{A}]-\frac{2}{3}\mathcal{A}\wedge[\mathrm{d}\mathcal{A}\wedge \mathcal{A}]}
\label{eq:A7}
\end{align}
which directly leads to \eqref{eq:A6} using $\braket{\mathcal{A}\wedge[\mathrm{d}\mathcal{A}\wedge \mathcal{A}]}=-\braket{\mathcal{A}\wedge[\mathcal{A}\wedge\mathrm{d}\mathcal{A}]}=-\braket{[\mathcal{A}\wedge \mathcal{A}]\wedge\mathrm{d}\mathcal{A}}$ which is an immediate consequence of identity \eqref{eq:A5}. When pulled back to the underlying bosonic submanifold $M$, the Chern-Simons action is thus defined as
\begin{equation}
S_{\mathrm{CS}}(\mathcal{A}):=\frac{k}{4\pi}\int_{M}{\braket{\mathcal{A}\wedge\mathrm{d}\mathcal{A}+\frac{1}{3}\mathcal{A}\wedge[\mathcal{A}\wedge\mathcal{A}]}}
\label{eq:A8}
\end{equation}
where $k$ is referred to as the \emph{level} of the Chern-Simons theory. Let us decompose $\mathcal{A}=\mathrm{pr}_{\mathfrak{g}_{\underline{0}}}\circ\mathcal{A}+\mathrm{pr}_{\mathfrak{g}_{\underline{1}}}\circ\mathcal{A}=:A+\psi$ w.r.t. the even and odd part of the super Lie algebra $\mathfrak{g}=\mathfrak{g}_{\underline{0}}\oplus\mathfrak{g}_{\underline{1}}$. Inserting this into \eqref{eq:A8}, this gives
\begin{equation}
\braket{\mathcal{A}\wedge F(\mathcal{A})}=\braket{A\wedge F(A)+\frac{1}{2}A\wedge[\psi\wedge\psi]}+\braket{\psi\wedge(\mathrm{d}\psi+[A\wedge\psi])}
\label{eq:A9}
\end{equation}
On the other hand, using $\braket{\psi\wedge[A\wedge\psi]}=\braket{\psi\wedge[\psi\wedge A]}=\braket{[\psi\wedge\psi]\wedge A}$ according to \eqref{eq:A5}, we find
\begin{align}
\braket{\mathcal{A}\wedge[\mathcal{A}\wedge\mathcal{A}]}&=\braket{A\wedge[A\wedge A]+A\wedge[\psi\wedge\psi]}+2\braket{\psi\wedge[A\wedge\psi]}\nonumber\\
&=\braket{A\wedge[A\wedge A]+A\wedge[\psi\wedge\psi]}+2\braket{A\wedge[\psi\wedge\psi]}\nonumber\\
&=\braket{A\wedge[A\wedge A]+3A\wedge[\psi\wedge\psi]}
\label{eq:A10}
\end{align}
Thus, we can rewrite \eqref{eq:A8} as follows
\begin{equation}
S_{\mathrm{CS}}(\mathcal{A})=S_{\mathrm{CS}}(A)+\frac{k}{4\pi}\int_{M}{\braket{\psi\wedge D^{(A)}\psi}}
\label{eq:A11}
\end{equation}
with $S_{\mathrm{CS}}(A)$ the Chern-Simons action of the bosonic connection $A$ and $D^{(A)}$ the associated exterior covariant derivative.

\section{The super Poncaré and anti-de Sitter group}\label{Appendix:Supergroups}
In this section, let us briefly review the basic supergroups and algebras that play a central role in context of supergravity in $D=4$ spacetime dimensions (see e.g. \cite{Nicolai:1984hb,Freedman:1983na,Freedman:2012zz,Wipf:2016} for a more detailed exposition as well as \cite{Eder:2022} for our choice of conventions).\\
Let $\gamma^I$, $I=0,\ldots,3$, be the gamma matrices satisfying the Clifford algebra relations $\left\{\gamma_{I},\gamma_{J}\right\}=2\eta_{IJ}$ with Minkowski metric $\eta$ with signature $\eta=\mathrm{diag}(-+++)$. We then define totally antisymmetric matrices $\Sigma^{AB}$, $A,B=0,\ldots,4$, via
\begin{equation}
\Sigma^{IJ}:=\frac{1}{2}\gamma^{IJ}:=\frac{1}{4}[\gamma^I,\gamma^J]\quad\text{as well as}\quad\Sigma^{4I}:=-\gamma^{I4}:=\frac{1}{2}\gamma^I
\label{eq:C1}
\end{equation}
where indices are raised and lowered w.r.t. the metric $\eta_{AB}=\mathrm{diag}(-+++-)$. These satisfy the following commutation relations
\begin{equation}
[\Sigma_{AB},\Sigma_{CD}]=\eta_{BC}\Sigma_{AD}-\eta_{AC}\Sigma_{BD}-\eta_{BD}\Sigma_{AC}+\eta_{AD}\Sigma_{BC}
\label{eq:C2}
\end{equation}
and thus provide a representation of $\mathfrak{so}(2,3)$, Lie algebra of the isometry group $\mathrm{SO}(2,3)$ of anti-de Sitter spacetime $\mathrm{AdS}_4$. Moreover, due to
\begin{equation}
(C\Sigma_{AB})^T=C\Sigma_{AB}
\label{eq:C3}
\end{equation}
with $C$ the charge conjugation matrix, it follows that $\Sigma_{AB}$ generate $\mathfrak{sp}(4)$ the Lie algebra universal covering group $\mathrm{Sp}(4,\mathbb{R})$ of $\mathrm{SO}(2,3)$\\
The graded extension of the anti-de Sitter group with $\mathcal{N}$-fermionic generators is given by the orthosymplectic Lie group $\mathrm{OSp}(\mathcal{N}|4)$ containing $\mathrm{O}(\mathcal{N})\times\mathrm{Sp}(4)$ as a bosonic subgroup and which, on the super vector space $\mathcal{V}=(\Lambda^{\mathbb{C}})^{\mathcal{N},4}$ with $\Lambda$ a real Grassmann-algebra, is defined w.r.t. the bilinear form induced by
\begin{equation}
\Omega=\begin{pmatrix}
	\mathds{1} & 0\\
	0 & C
\end{pmatrix}
\label{eq:C4}
\end{equation}
The algebra $\mathfrak{osp}(\mathcal{N}|4)$ is then generated by all $X\in\mathfrak{gl}(\mathcal{V})$ satisfying
\begin{equation}
X^{sT}\Omega+\Omega X=0
\label{eq:C5}
\end{equation}
where $X^{sT}$ denotes the super transpose of $X$. The bosonic generators of super Lie algebra are given by
\begin{equation}
M_{AB}:=\begin{pmatrix}
	0 & 0\\
	0 & \Sigma_{AB}
\end{pmatrix}\quad\text{and}\quad T^{rs}:=\begin{pmatrix}
	A^{rs} & 0\\
	0 & 0
\end{pmatrix}
\label{eq:C6}
\end{equation}
respectively, where $(A^{rs})_{pq}:=2\delta_p^{[r}\delta_q^{s]}$, $p,q,r,s=1,\ldots,\mathcal{N}$. The fermionic generators are given by
\begin{equation}
Q_{\alpha}^r:=\begin{pmatrix}
	0 & -\bar{e}_{\alpha}\otimes e_r\\
	e_{\alpha}\otimes e_r^T & 0
\end{pmatrix}
\label{eq:C7}
\end{equation}
with $(\bar{e}_{\alpha})_{\beta}=C_{\alpha\beta}$. Setting $P_I:=\frac{1}{L}\Sigma_{4I}$, and rescaling $Q_{\alpha}^r\rightarrow Q_{\alpha}^r/\sqrt{2L}$ as well as $T^{rs}\rightarrow T^{rs}/2L$, one obtains the following (graded) commutation relations
\begin{align}
[M_{IJ},Q^r_{\alpha}]&=\frac{1}{2}Q_{\beta}^r\tensor{(\gamma_{IJ})}{^{\beta}_{\alpha}}\label{eq:C8a}\\
[P_I,Q^r_{\alpha}]&=-\frac{1}{2L}Q^r_{\beta}\tensor{(\gamma_I)}{^{\beta}_{\alpha}}\label{eq:C8b}\\
[T^{pq},Q_{\alpha}^r]&=\frac{1}{2L}(\delta^{qr}Q_{\alpha}^p-\delta^{pr}Q_{\alpha}^q)\label{eq:C8c}\\
[Q_{\alpha}^r,Q_{\beta}^s]=\delta^{rs}\frac{1}{2}(C\gamma^I)_{\alpha\beta}P_I+&\delta^{rs}\frac{1}{4L}(C\gamma^{IJ})_{\alpha\beta}M_{IJ}-C_{\alpha\beta}T^{rs}
\label{eq:C8d}
\end{align}
which in the limit $L\rightarrow\infty$ leads to the respective super Poincaré Lie algebra.\\
The orthosymplectic and Poincaré superalgebra contain a proper subalgebra which appears in context of chiral supergravity.  Let $T^{\pm}_i$ be defined as  
\begin{equation}
T_i^{\pm}=\frac{1}{2}(-\frac{1}{2}\tensor{\epsilon}{_{i}^{jk}}M_{jk}\pm iM_{0i})
\label{eq:C9}
\end{equation}
satisfying the commutation relations 
\begin{equation}
[T^{\pm}_i,T_j^{\pm}]=\tensor{\epsilon}{_{ij}^k}T_k^{\pm}
\label{eq:C10}
\end{equation} 
Since, the $R$-symmetry generators do not mix the chiral components of the Majorana generators $Q_{\alpha}^r$, it follows that $(T_i^+,T_{rs},Q_{A}^r)$ form a proper chiral sub super Lie algebra of $\mathfrak{osp}(\mathcal{N}|4)_{\mathbb{C}}$ with the graded commutation relations
\begin{align}
[T^+_i,T_j^+]&=\tensor{\epsilon}{_{ij}^k}T_k^+\label{eq:C11a}\\
[T_i^+,Q^r_A]&=Q^r_{B}\tensor{(\tau_i)}{^B_A}\label{eq:C11b}\\
[Q^r_{A},Q^s_{B}]&=\delta^{rs}\frac{1}{L}(\epsilon\sigma^i)_{AB}T_i^+-\frac{i}{2L}\epsilon_{AB}T^{rs}\label{eq:C11c}\\
[T^{pq},Q_{A}^{r}]&=\frac{1}{2L}(\delta^{qr}Q_{A}^p-\delta^{pr}Q_{A}^q)
\label{eq:C11d}
\end{align}
yielding the complex orthosymplectic Lie superalgebra $\mathfrak{osp}(\mathcal{N}|2)_{\mathbb{C}}$, the extended supersymmetric generalization of the isometry algebra of $D=2$ anti-de Sitter space . In the limit $L\rightarrow\infty$, this yields the extended $D=2$ super Poincaré algebra.

\end{document}